\documentclass[twocolumn]{aastex61}

\usepackage{multirow}

\usepackage[english]{babel}
\usepackage[utf8x]{inputenc}
\usepackage[T1]{fontenc}

\usepackage{amsmath}
\usepackage{graphicx}
\begin{document}

\title[Probing actinide production under different conditions]{Probing the Production of Actinides Under Different r-Process Conditions}

\author[0000-0002-4445-8908]{M Eichler}
\affiliation{Institut f\"ur Kernphysik, Technische Universit\"at Darmstadt, Schlossgartenstr. 2, DE-64289 Darmstadt, Germany}
\author{W Sayar}
\affiliation{Institut f\"ur Kernphysik, Technische Universit\"at Darmstadt, Schlossgartenstr. 2, DE-64289 Darmstadt, Germany}
\author{A Arcones}
\affiliation{Institut f\"ur Kernphysik, Technische Universit\"at Darmstadt, Schlossgartenstr. 2, DE-64289 Darmstadt, Germany}
\affiliation{GSI Helmholtzzentrum f\"ur Schwerionenforschung GmbH, Planckstr. 1, DE-64291 Darmstadt, Germany}
\author[0000-0002-1266-0642]{T Rauscher}
\affiliation{Department of Physics, University of Basel, Klingelbergstrasse 82, 4056 Basel, Switzerland}
\affiliation{Centre for Astrophysics Research, University of Hertfordshire, College Lane, Hatfield AL10 9AB, United Kingdom}

\submitjournal{ApJ}

\begin{abstract}
Several extremely metal-poor stars are known to have an enhanced thorium abundance. These actinide-boost stars have likely inherited material from an r-process that operated under different conditions than the r-process that is reflected in most other metal-poor stars with no actinide enhancement.

In this article, we explore the sensitivity of actinide production in r-process calculations on the hydrodynamical conditions as well as on the nuclear physics. We find that the initial electron fraction $Y_e$ is the most important factor determining the actinide yields and that the abundance ratios between long-lived actinides and lanthanides like europium can vary for different conditions in our calculations. In our setup, conditions with high entropies systematically lead to lower actinide abundances relative to other r-process elements. Furthermore, actinide-enhanced ejecta can be distinguished from the ``regular'' composition also in other ways, most notably in the second r-process peak abundances.

\end{abstract}

\keywords{nucleosynthesis, actinide-boost stars, r-process, kilonova}

\section{Introduction}

The \textit{rapid neutron capture process} (r-process) produces about half of the elements heavier than iron in the universe \citep{bbfh1957}. All of the observable actinide abundances originate from the r-process, since other nucleosynthesis processes synthesizing heavy nuclei cannot overcome the unstable elements beyond Bi to form the long-lived actinides $^{232}$Th ($T_{1/2} = 14.0$~Gyr), $^{235}$U ($T_{1/2} = 0.7$~Gyr), and $^{238}$U ($T_{1/2} = 4.5$~Gyr). In particular, a) the s-process is terminated by the $\alpha$-decay of $^{210}$Po, b) the $\gamma$-process photodissociates heavy (mostly s-process) seed nuclei, and c) the $\nu$p-process and explosive $\alpha$-captures only reach moderate mass numbers of around 120, because the high temperatures required for the charged-particle reactions also facilitate the reverse (photodissociation) reactions. At the same time, the long half-lives mean that these isotopes can be used for age determinations of metal-poor stars (e.g., \citealt{cayrel2001,hill2002,schatz2002,frebel2007,roederer2009}).

The recent groundbreaking observation of gravitational waves from a neutron star merger (GW170817) and the subsequent electromagnetic signal originating from decaying r-process nuclei (kilonova or macronova; AT~2017gfo or
SSS17a) confirmed neutron star mergers as an r-process site \citep{GW170817-EM,Cowperthwaite2017,kasliwal2017,kilpatrick2017,metzger2017,pian2017a,tanaka2017,tanvir2017,rosswog2018}. 
The characteristics of the light curve reveals that the ejecta can be divided into at least two components: (a) low-$Y_e$ ejecta ($Y_e < 0.3$) responsible for the production of lanthanides and possibly actinides, and (b) material with a higher $Y_e$ that contained only a small fraction of lanthanides at most. What makes this distinction possible is the very high opacity of lanthanides and actinides \citep{kasen2013}, which means that lanthanide-rich environments become transparent to $\gamma$-rays only after one to several days. 
The upper limit of $Y_e \approx 0.3$ to produce a significant amount of lanthanides has been established by different independent studies (e.g., \citealt{wanajo2014,lippuner2015,goriely2015b,rosswog2018}). The inferred properties of the observed kilonova leave space for different interpretations. 
For instance, \cite{rosswog2018} and \cite{wanajo2018} demonstrate that the light curve at late times (1-10~days) could be dominated by $\beta$-decays of lighter nuclei. \cite{wanajo2018} have identified two $\beta$-decay chains of light trans-iron nuclei as possible main contributors to the luminosity a few days after the event: $^{66}$Ni~$\rightarrow ^{66}$Cu~$\rightarrow ^{66}$Zn and $^{72}$Zn~$\rightarrow ^{72}$Ga~$\rightarrow ^{72}$Ge, both with half-lives around 2~days (also discussed in \citealt{wu2019}).

The simultaneous detection of GW170817 and \linebreak AT~2017gfo represents the first-ever direct observation of r-process-rich ejecta in an astrophysical environment. However, despite this breakthrough open questions remain. If neutron star mergers are assumed to be the only r-process site in the universe, discrepancies arise in models of galactic chemical evolution (GCE) and cosmological zoom simulations that use realistic delay-time distributions for NSMs \citep{cote2017,hotokezaka2018,cote2018,safarzadeh2018,simonetti2019}. Furthermore, Eu-enriched ultra-metal-poor stars are generally hard to reconcile with the neutron star merger scenario, since two core-collapse supernovae are needed in order to produce the neutron stars, thus heavily polluting the neighbourhood with iron long before the merger can produce europium \citep{wehmeyer2015,cote2018,HaynesKobayashi2019}. The heaviest r-process nuclei can potentially be produced in several scenarios. In this article, we will focus on three sites: (a) prompt (dynamical) ejecta, (b) late-time disk ejecta in neutron star mergers, and (c) magneto-hydrodynamically driven supernovae (MHD SNe). 
Even after GW170817, an additional r-process site could still be needed to explain observed trends in galactic chemical evolution (see, e.g., \citealt{cote2018}).

It has been established that the abundances of r-process elements in r-process enhanced metal-poor stars follow the solar (residual) r-process composition remarkably well \citep{sneden2008}. However, since the attempts by \cite{hill2002} and \cite{schatz2002} to apply the nucleochronometers Th and U to the star CS~31082-001, more and more stars are found to have an enhanced actinide abundance (although most of the time only Th can be measured), in comparison to other stars and nucleosynthesis models. Even before these observations, theoretical calculations have shown that actinide production in the r-process can vary depending on the conditions and the nuclear mass model (e.g., \citealt{goriely1999}). An overview on actinide-boost stars has been given in \cite{roederer2009}, with more recent discoveries by \cite{mashonkina2014} and \cite{holmbeck2018a}. \cite{ji2018}, on the other hand, have measured the Th abundance of DES~J033523−540407, a star in the r-process-enriched ultra-faint dwarf galaxy Reticulum~II, and their result suggests that this star might belong to a different, actinide-deficient category of stars. These recent observations raise the question (also discussed, e.g., in \citealt{holmbeck2019b}): Is the r-process event that is responsible for the actinide boost the same as the one without actinide-boost, but with different conditions in the ejecta (i.e., most likely neutron star mergers), or is the variation in actinide content a sign of more than one r-process site in the universe?

In the present study, we explore the sensitivity of actinide production in r-process calculations on the hydrodynamical conditions as well as on the nuclear physics. To that end, we employ and compare six hydrodynamical models of three possible r-process sites, three nuclear mass models, and two sets of theoretical $\beta$-decay rates, thus establishing the dependence of actinide yields on a wide range of hydrodynamical and nuclear conditions. Furthermore, we discuss possibilities how actinide-boosted ejecta can be distinguished in future kilonova observations as well as potential deviations from the robust r-process abundance pattern in actinide-boost stars that have yet to be tested. 

This paper is structured as follows: Section~\ref{sec:method} describes the nuclear network and the hydrodynamical models used for this study. Sections~\ref{sec:results}~\&~\ref{sec:discussion} present and analyze the results. Possible observational signatures of actinide-rich environments are discussed in section~\ref{sec:observables}, followed by our conclusions in section~\ref{sec:conclusions}.

\section{Method}
\label{sec:method}
\subsection{Nuclear Network}

\begin{table*}
\caption{Overview of reaction rates used for this study.\label{tab:reactionrates}}
\begin{tabular}{llll}
\multicolumn{1}{c}{Library} & \multicolumn{1}{c}{(n,$\gamma$), ($\gamma$,n)} & \multicolumn{1}{c}{fission} & \multicolumn{1}{c}{$\beta$-decay} \\
\hline
FRDM & \cite{rauscher2000} & \cite{panov2010} & \cite{moeller2003} \\
 & (FRDM set, $Z\leq 83$) & (TF barriers, neutron-induced) &  \\
 & \cite{panov2010} & \cite{panov2005} & \\
 & (FRDM set, $Z>83$) & (TF barriers, $\beta$-delayed) & \\
 & & \cite{petermann2012} & \\
 & & (TF barriers, spontaneous) & \\
ETFSI-Q & \cite{rauscher2000} & \cite{panov2010} & \cite{moeller2003} \\
 & (ETFSI-Q set, $Z\leq 83$) & (ETFSI-Q barriers, neutron-induced) &  \\
 & \cite{panov2010} & \cite{panov2005} & \\
 & (ETFSI-Q set, $Z>83$) & (ETFSI-Q barriers, $\beta$-delayed) & \\
 & & \cite{panov2013} & \\
 & & (ETFSI-Q barriers, spontaneous) & \\
DZ10 & SMARAGD calculation & \cite{panov2010} & \cite{moeller2003} \\
 & ($Z\leq 83$) & (TF barriers, neutron-induced) &  \\
 & \cite{panov2010} & \cite{panov2005} & \\
 & (FRDM set, $Z>83$) & (TF barriers, $\beta$-delayed) & \\
 & & \cite{petermann2012} & \\
 & & (TF barriers, spontaneous) & \\
D3C*(FRDM) & \cite{rauscher2000} & \cite{panov2010} & \cite{marketin2016} \\
 & (FRDM set, $Z\leq 83$) & (TF barriers, neutron-induced) & \\
 & \cite{panov2010} & \cite{panov2005} & \\
& (FRDM set, $Z>83$) & (TF barriers, $\beta$-delayed)* & \\
 & & \cite{petermann2012} & \\
 & & (TF barriers, spontaneous) & \\
\hline
\multicolumn{4}{l}{*$\beta$-delayed fission rates are given as a fraction of the total $\beta$-decay rate, so the $\beta$-delayed fission rates in} \\ 
\multicolumn{4}{l}{ D3C*(FRDM) and FRDM differ, although they are based on the same barriers.}
\end{tabular}
\end{table*}

We perform our nucleosynthesis calculations using four different libraries of nuclear reactions to run our nuclear network code \textsc{Winnet} \citep{winteler2012}. The first set is the JINA Reaclib default version (from 10/20/2017), with added rates for neutron-induced fission from \cite{panov2010}, $\beta$-delayed fission from \cite{panov2005}, and spontaneous fission as described in \cite{petermann2012} (henceforth referred to as FRDM). Note that this JINA Reaclib set includes theoretical rates from the FRDM set of \cite{rauscher2000} on the neutron-rich side and the theoretical neutron capture and ($\gamma$,n) rates from \cite{panov2010} for (n,$\gamma$) target nuclei with $Z>83$. In addition to this reaction library, we also perform calculations based on the Extended Thomas-Fermi with Strutinsky Integral mass model including shell quenching corrections, ETFSI-Q \citep{aboussir1995}. In this set, the neutron capture rates and their reverse reactions are taken from the ETFSI-Q rate set given in \cite{rauscher2000}, supplemented by neutron capture rates on $Z>83$ nuclei from \cite{panov2010}, spontaneous fission rates from \cite{panov2013}, and the rates for the other fission modes from the same sources as above, but based on the corresponding fission barriers \citep{mamdouh2001}. A third reaction library consists of rates based on the Duflo-Zuker mass model \citep{duflozuker1995} with 10 parameters (labeled as DZ10 in the following), newly calculated for this work using the SMARAGD Hauser-Feshbach code as described in \citep{rauscher2011,cyburt2010}. A fourth set is a variation of the first library, but with theoretical $\beta$-decay half-lives replaced with the predictions of \cite{marketin2016}. This library will be referred to as D3C*(FRDM) throughout this article. Table \ref{tab:reactionrates} provides an overview of the reaction rates in the different libraries that are relevant to this study.

Nuclear reactions release energy which can increase the temperature. We include this effect in our nucleosynthesis calculations following the description of \cite{freiburghaus1999}.

\subsection{Hydrodynamical Models}
In order to test actinide production in the r-process in different conditions, we employ several models of suggested r-process sites. We are using simulations of dynamical ejecta in a binary compact merger (two neutron stars with 1.0~M$_{\odot}$ each; in the following called R1010) and a neutron star - black hole merger (1.4~M$_{\odot}$ and 5.0~M$_{\odot}$; henceforth referred to as R1450) from \cite{korobkin2012, rosswog2013}. We also include a neutron star merger model from \cite{bovard2017}, with both neutron star masses of 1.25~M$_{\odot}$ and the SFHO equation of state \citep{steiner2013}. As discussed in \cite{bovard2017}, the dynamical ejecta in their models cover a wider range of $Y_e$ and entropies than the \cite{rosswog2013} models, although the bulk of the ejecta also contains rather low electron fractions and entropies. Moreover, we test other possible sites of the r-process: accretion disks in mergers have been shown to host conditions favourable for a strong r-process \citep{surman2008, fernandez2013b, perego2014, just2015, wu2016, lippuner2017, SiegelMetzger2018}. Therefore, we include two different disk scenarios (\textit{S-def} and \textit{S-s6}) which were first described in \cite{fernandez2013b}. The data we are using are from the improved simulations as described in \cite{wu2016, lippuner2017}. Furthermore, magneto-hydrodynamically driven (MHD) supernovae with fast expanding jets possibly represent an additional r-process site, where heavy elements could be produced under different conditions than in compact binary mergers. Here we are using the model of \cite{winteler2012}. Other MHD SN simulations have been performed by, e.g., \cite{nishimura2015,nishimura2017,moesta2018}. As a summary, all hydrodynamical models are listed in table~\ref{tab:models}, together with the nomenclature that will be used throughout the text.

The hydrodynamic trajectories provide data only up to the end of the simulation, which is typically of the order of 0.01~s for explosive models (and a few seconds for the disk models). We therefore extrapolate using a parametrized expansion according to \cite{korobkin2012} and \cite{eichler2015}. If not indicated otherwise, the results shown represent abundances 1~Gyr after the r-process event.

\begin{table*}
\begin{center}
\caption{Overview of hydrodynamical models used for this study. 
\label{tab:models}}
\begin{tabular}{ccccc}
type & masses [M$_{\odot}$] & reference & name in reference & model name here \\
\hline
NS-NS dyn. ejecta & 1.0 \& 1.0   & \cite{korobkin2012} & Run 1 & R1010\\
NS-BH dyn. ejecta & 1.4 \& 5.0   & \cite{korobkin2012} & Run 22 & R1450\\
NS-NS dyn. ejecta & 1.25 \& 1.25 & \cite{bovard2017}  & SFHO-M1.25 & Bs125\\
NS-NS disk        & 3.0 \& 0.03  & \cite{wu2016} & S-def & FMdef\\
NS-NS disk        & 3.0 \& 0.03  & \cite{wu2016} & S-s6  & FMs6\\
MHD SN            & 15.0         & \cite{winteler2012}   & w/ $\nu$ heating & Wmhd\\ 
\hline
\end{tabular}
\end{center}
\end{table*}

\section{Results}
\label{sec:results}

\subsection{Production channels of $^{232}$Th}
\begin{figure}[t]
\includegraphics[width=\columnwidth]{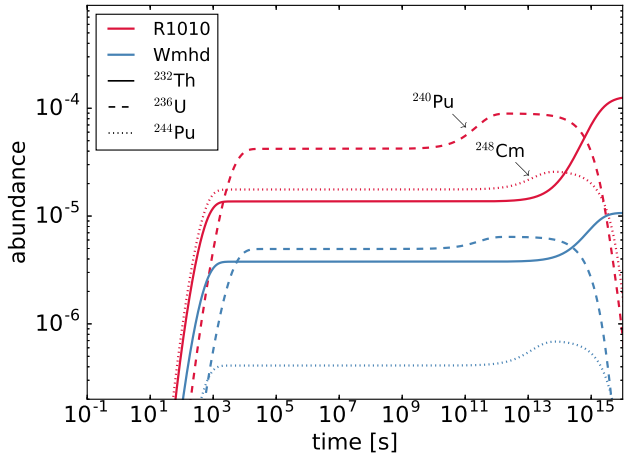}
\caption{Evolution of $^{232}$Th, $^{236}$U, and $^{244}$Pu for two r-process calculations using one trajectory from R1010 (red) and from Wmhd (blue), respectively. \label{fig:actevol}}
\end{figure}

The main goal of our study is to investigate and explain trends of $^{232}$Th (actinide) production for different conditions and theoretical nuclear physics models. For this purpose, we summarize here how $^{232}$Th is built up by decaying r-process nuclei on the example of a trajectory from R1010 and one from Wmhd.  In Figure~\ref{fig:actevol} the abundances of $^{232}$Th, $^{236}$U, and $^{244}$Pu are shown as a function of time. Like for any other (quasi-)stable isotope produced in the r-process, there is a direct $\beta$-decay feeding channel from more neutron-rich nuclei with mass numbers A $\gtrsim 232$ (including the possibility of $\beta$-delayed neutron emission), responsible for the initial strong build-up around 100~s (see Fig.~\ref{fig:actevol}). An additional production channel only sets in much later, when $^{236}$U begins to $\alpha$-decay with a half-life of $23.4$~Myr. Before it decays, $^{236}$U itself is fed by the $\alpha$-decay of $^{240}$Pu, which in turn is added to by the decay chain $ ^{248}\textrm{Cm}(\alpha \gamma)^{244}\textrm{Pu}(\alpha \gamma)^{240}\textrm{U}(\beta^-\bar{\nu})^{240}\textrm{Np}(\beta^-\bar{\nu})^{240} \textrm{Pu}$. Note that all of these isotopes are also directly produced by the $\beta$-decay channel. The $^{232}$Th abundance reaches its maximum value only after $^{244}$Pu has decayed, with a half-life of $80.0$~Myr. Figure~\ref{fig:actevol} furthermore reveals that a difference in $^{232}$Th abundances can have different origins. While the Wmhd case produces less actinides in general (resulting in a lower $^{232}$Th production after 100~s compared to the R1010 case), the largest difference is in the heavier isotopes $^{240,244}$Pu. Thus, the initial difference further increases when the heavier actinides $\alpha$-decay around the $10^{15}$~s mark.

The plurality of production channels effectively means that nuclei with a wide mass number range ($232 \lesssim A \lesssim 250$) are responsible for the eventual production of $^{232}$Th. This mass range coincides with relatively long-lived nuclei on the r-process path along the $N=162$ isotone. In particular, $^{242}$Hg and $^{241}$Au are strongly produced in our calculations, acting as important precursor nuclei for the final $^{232}$Th abundance. The dependence of our results on the $\beta$-decay half-lives of these two nuclei is discussed in section~\ref{sec:discuss_nuc}. 

\begin{figure*}
  \includegraphics[width=\textwidth]{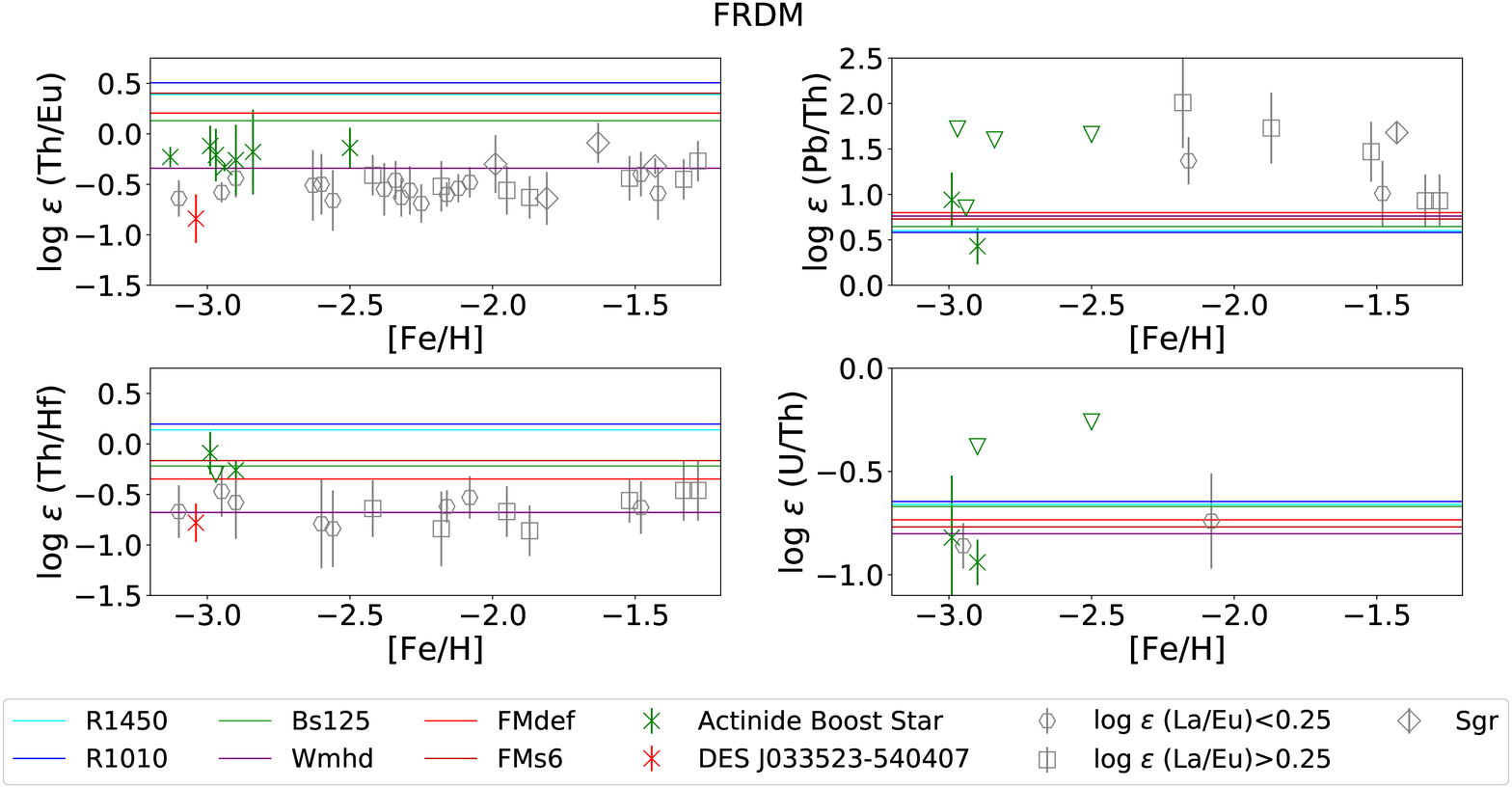} \\
  \includegraphics[width=\textwidth]{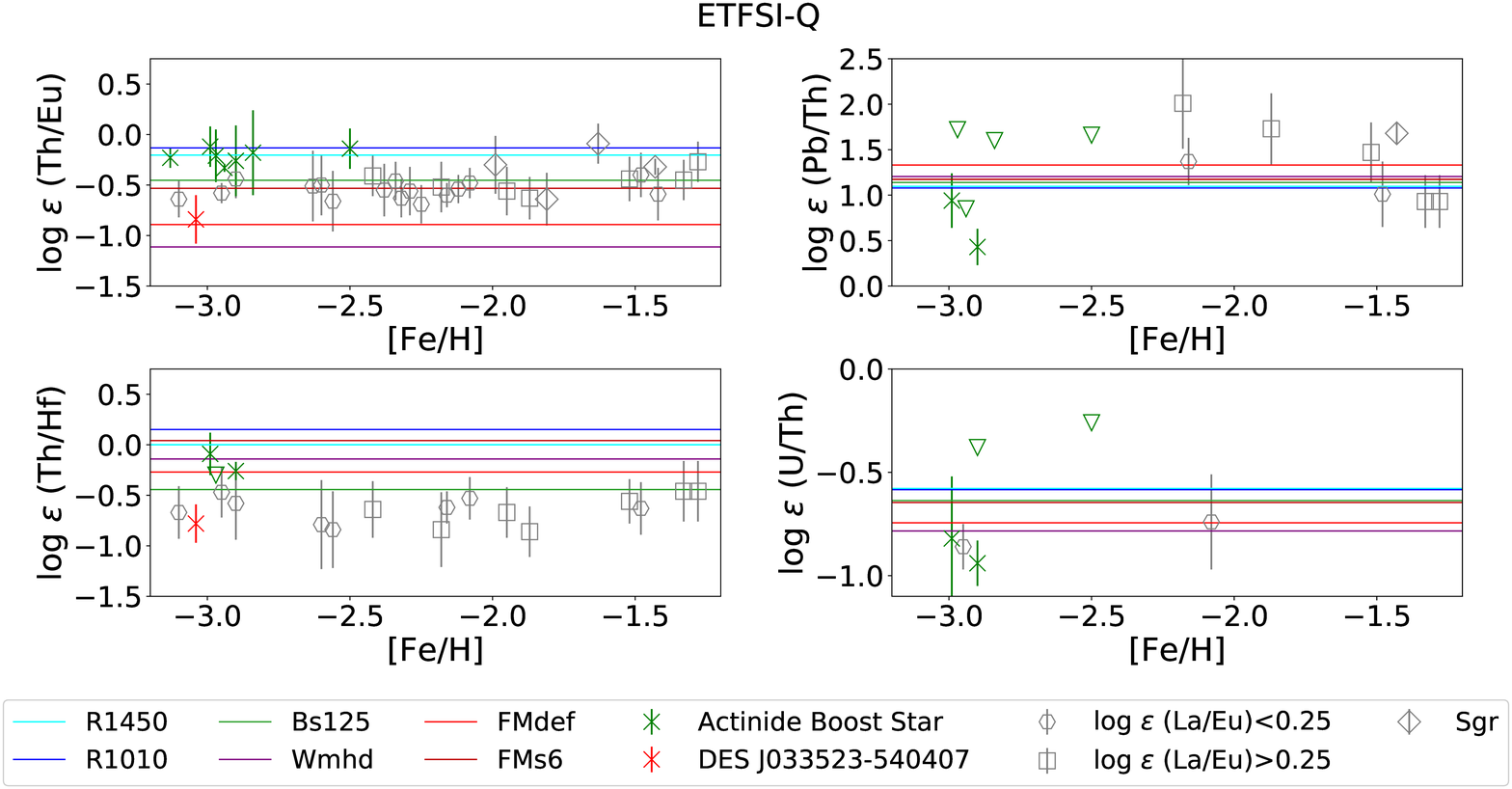}
  \caption{Abundance ratios of chronometer pairs Th/Eu (top left), Th/Hf (bottom left), Pb/Th (top right), and U/Th (bottom right). In each panel theoretical abundance ratios 10 Gyr after the event are shown for different hydrodynamical models of r-process sites as horizontal lines, and compared to stellar observations. The theoretical reaction rate libraries used here are FRDM (top) and ETFSI-Q (bottom). The down-facing triangles indicate upper limits. See text for further details. \label{fig:observ_ratios}}
\end{figure*}

\begin{figure*}
  \includegraphics[width=\textwidth]{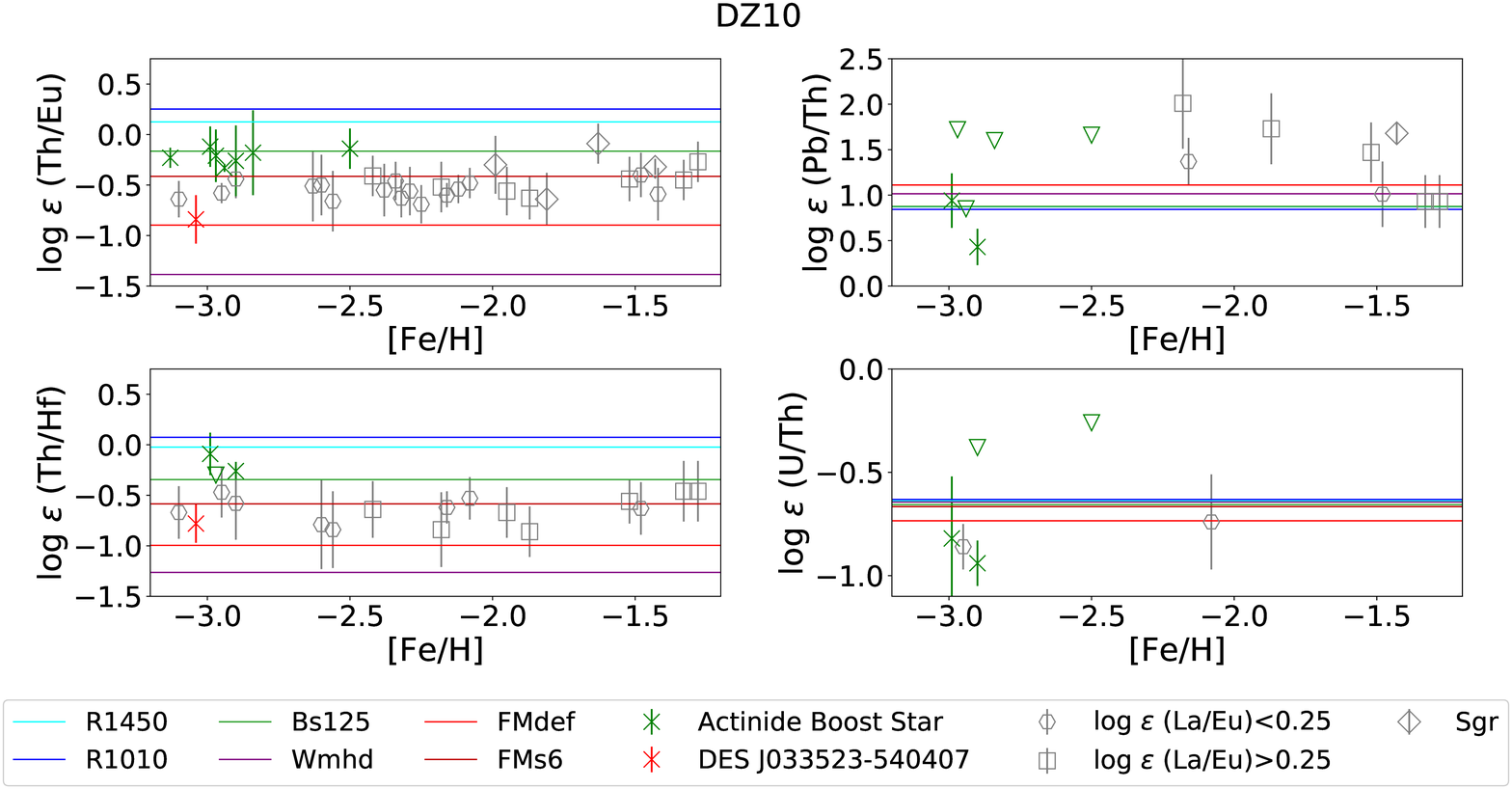} \\
  \includegraphics[width=\textwidth]{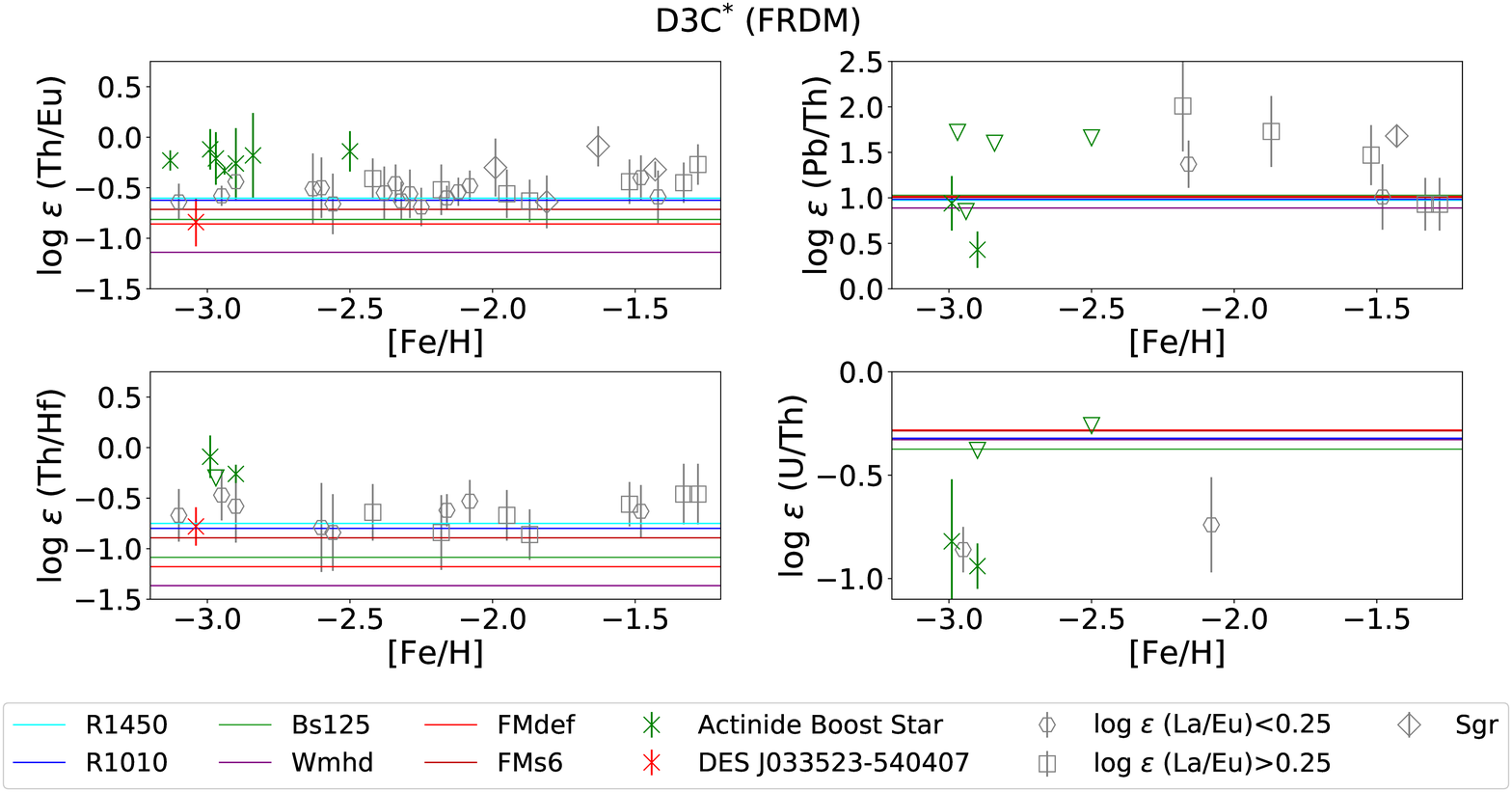}
  \caption{Same as Fig.~\ref{fig:observ_ratios}, but for the DZ10 and D3C*(FRDM) reaction libraries. \label{fig:observ_ratios2}}
\end{figure*}

\subsection{Actinide and lanthanide yields in our models}

Here we want to study the impact of astrophysical and nuclear factors on the actinide production in r-process calculations. The models described in section~\ref{sec:method} all produce the heaviest r-process nuclei up to the actinides. However, the conditions in the ejecta differ considerably (i.e., electron fraction, entropy, and ejecta velocity), which leaves an imprint on the compositions. \cite{roederer2009} have tested their sample of stellar abundance ratios against a site-independent nucleosynthesis model consisting of several components with different neutron densities and employing the ETFSI-Q mass model. In Figures~\ref{fig:observ_ratios}~\&~\ref{fig:observ_ratios2} we revisit their approach with real hydrodynamical r-process models for all our four sets of nuclear physics input. The four panels belonging to each reaction rate library show four different chronometer pairs: Th/Eu (top left), Th/Hf (bottom left), Pb/Th (top right), and U/Th (bottom right). The dots represent stellar abundance data from \cite{roederer2009}, complemented by the newer sample of \cite{mashonkina2014}, as well as the recent survey of stars in the Sagittarius dwarf galaxy by \cite{hansen2018} and the recently discovered actinide-boost star J09544277+5246414 \citep{holmbeck2018a}. In addition, we also include the Reticulum II star with known Th yield, DESJ033523-540407 \citep{ji2018}. In accordance with \cite{roederer2009}, stars with $\log_{\epsilon} \left( \textrm{La/Eu} \right) < 0.25$ are marked separately, since they are assumed to have no significant contribution from the s-process to the heavy element yields. Actinide-boost stars are marked in green. 
The calculated abundance ratios are shown in Figures~\ref{fig:observ_ratios}~\&~\ref{fig:observ_ratios2} as horizontal lines for all hydrodynamical models (see table~\ref{tab:models}) for a time $t=10$~Gyr after the event.

Figures~\ref{fig:observ_ratios}~\&~\ref{fig:observ_ratios2} reveal the dependence of the chronometer pairs on the nuclear physics as well as the hydrodynamical conditions of the r-process. Scenarios with a very low $Y_e$ generally contain the highest Th/Eu and Th/Hf abundance ratios (i.e., R1010 and R1450), while the more moderate electron fractions prevalent in the merger disks and MHD SN result in lower ratios. An exception from this general trend can be observed in the Th/Hf ratio with the ETFSI-Q mass model, where the Bs125 model produces the lowest abundance ratio. Across the nuclear mass models, FRDM exhibits the highest Th yields, visible here from the highest Th/Eu and Th/Hf and the lowest Pb/Th ratios of all sets considered. The ETFSI-Q model, on the other hand, has the lowest Th/Eu ratio for the merger scenarios. The DZ10 model shows the largest sensitivity on the hydrodynamical scenarios, with theoretical Th/Eu and Th/Hf ratios spanning one and a half orders of magnitude (compared to less than one order of magnitude for both FRDM and ETFSI-Q). The Pb/Th and U/Th abundance ratios are quite insensitive to the hydrodynamical models, suggesting that these chronometer pairs are more reliable than Th/Eu for the determination of stellar ages. With regards to the nuclear physics, Pb/Th and U/Th are also less sensitive than Th/Eu and Th/Hf. A comparison of the D3C* set with FRDM illustrates the impact of the $\beta$-decay rates. As already mentioned, the D3C* half-lives are systematically shorter for heavy nuclei, which results in lower actinide abundances (see also \citealt{eichler2015,wu2016,holmbeck2019a}) and thus significantly lower Th/Eu and Th/Hf ratios. However, Pb/Th and U/Th increase for all hydrodynamical models compared to the FRDM calculations. Since the thorium-to-lanthanide abundance ratios are highest for R1010, R1450, and Bs125 (i.e., models of NSM dynamical ejecta), one possibility is that actinide-boost stars have inherited an r-process composition reflecting a larger fraction of dynamical ejecta and a smaller fraction of disk ejecta, compared to stars with a thorium-to-lanthanide abundance ratio closer to the solar value.

\begin{table*}
\begin{center}
\caption{Representative trajectories for each model employed. The second column lists the Th/Eu abundance ratio calculated with the original trajectory and evaluated at $t=1$~Gy after the event. The electron fraction $Y_e$, entropy $S$, and expansion velocity $v$ are read from the trajectory where the temperature drops below 8~GK for the last time. 
\label{tab:trajectories}}
\begin{tabular}{ccccccc}
name & $\log \epsilon$ (Th/Eu) & $Y_e$ (T$ = 8$~GK) & $S$ & $v$ & Ref. \\
 &   &  & $[k_B/\textrm{baryon}]$ & [cm/s] & \\
\hline
R1010-rep & 0.619 & 0.044 & 0.012 & 3.06 $\times 10^{9}$ & \cite{rosswog2013} \\
R1450-rep & 0.766 & 0.016 & 1.798 & 1.65 $\times 10^{9}$ & \cite{rosswog2013} \\
Bs125-rep & 0.337 & 0.225 & 37.367 & 7.87 $\times 10^9$ & \cite{bovard2017} \\
Wmhd-rep & -0.085 & 0.175 & 6.001 & 2.02 $\times 10^9$ & \cite{winteler2012} \\
FMdef-rep & 0.857 & 0.168 & 12.593 & 1.70 $\times 10^8$ & \cite{fernandez2013b} \\
FMs6-rep & 0.052 & 0.175 & 12.696 & 6.02 $\times 10^8$ & \cite{fernandez2013b} \\
\hline
\end{tabular}
\end{center}
\end{table*}

\subsection{Hydrodynamical conditions}

Typically, r-process conditions are described by three defining quantities: electron fraction $Y_e$, entropy $S$, and the expansion velocity, which describes how fast the ejecta expand and cool. In order to study the dependence of actinide production on the hydrodynamic properties of the r-process environment, we want to cover a wide range of different conditions that are actually present in models of possible r-process sites. To that end, we pick representative trajectories from our hydrodynamical models, i.e., one trajectory with conditions that are characteristic for each model. Their properties (electron fraction $Y_e$, entropy $S$, and expansion velocity $v$) are summarized in table~\ref{tab:trajectories}, along with the calculated Th/Eu abundance ratios for each trajectory.

\begin{figure*}
  \includegraphics[width=\textwidth]{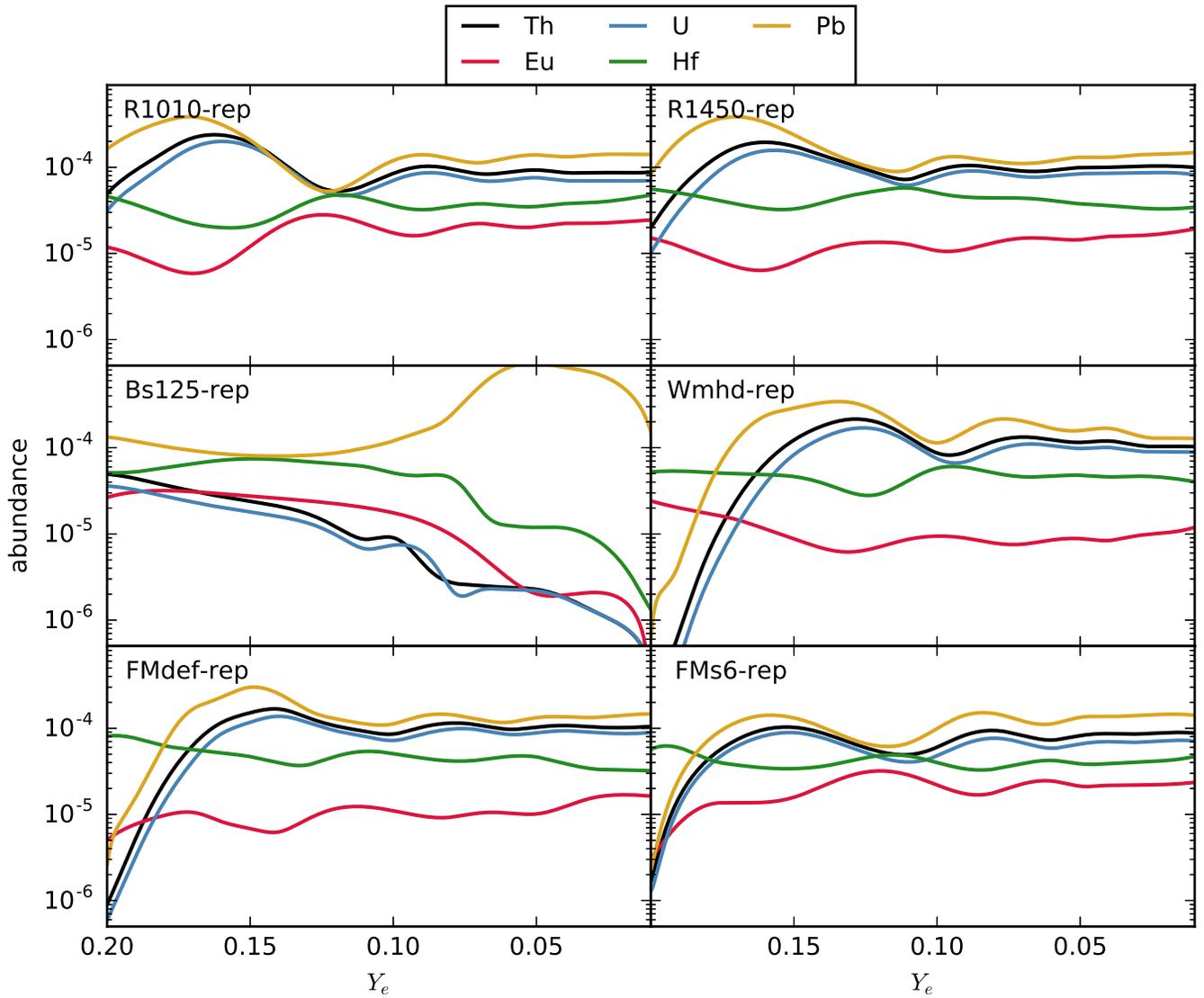} 
  \caption{\label{fig:Th/Eu_trends} Abundances of Th, U, Pb, Eu, and Hf at $t=1$~Gy after the r-process event for our six representative trajectories (see Table~\ref{tab:trajectories}) in dependence of initial $Y_e$. The mass model used for these calculations is FRDM.}
\end{figure*}

Using these representative trajectories, we now vary the initial $Y_e$ and repeat the nucleosynthesis calculations, tracking the actinide production as well as second peak and rare-earth peak elements, similar to the procedure described in \cite{holmbeck2019a}. The Th, U, Pb, Hf, and Eu abundances in dependence of initial $Y_e$ are shown in Figure~\ref{fig:Th/Eu_trends}. For R1010-rep the Th, U, and Pb yields have a distinct non-linear dependence on $Y_e$, with the global maximum around $Y_e \approx 0.16$ and a minimum at $Y_e \approx 0.12$ (comparable to the results of \citealt{holmbeck2019a}), while Eu exhibits the exact opposite trend. The NS-BH trajectory R1450-rep shows the same trend (since it has almost the same initial conditions, see table~\ref{tab:trajectories}). For Fmdef-rep, FMs6-rep, and Wmhd-rep the trend is shifted towards lower $Y_e$ values, with the strongest Th abundance peak below $Y_e=0.15$ and a minimum around $Y_e=0.1$. Only Bs125-rep shows a different trend, with an increasing Pb abundance towards lower $Y_e$, at the expense of all other four elements shown here. The peculiar composition at low $Y_e$ in this trajectory is the result of an r-process with very few seed nuclei and a fast expansion, leading to an extreme shift of the r-process peaks due to late neutron captures while the composition is decaying towards stability. This means that under these conditions the third peak composition is effectively dominated by lead nuclei.

\begin{figure}
  \includegraphics[width=0.5\textwidth]{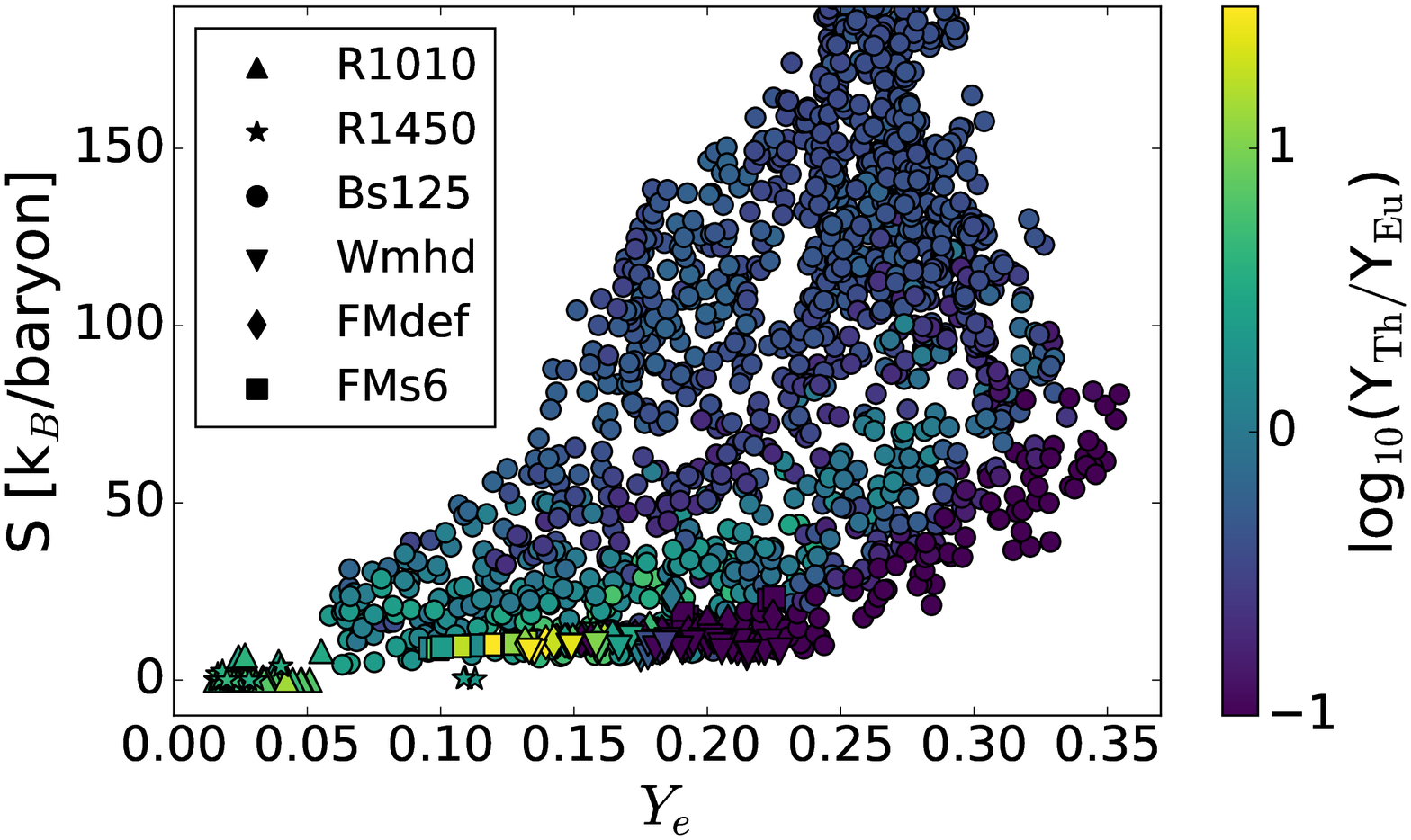}
  \caption{Th/Eu abundance ratio ($t=1$~Gyr) for individual trajectories in dependence of entropy $S$ and electron fraction $Y_e$ at the time when the temperature drops below 8~GK for the last time in the trajectory. The symbols identify the hydrodynamical model each trajectory belongs to. \label{fig:ThEu_univ}}
\end{figure}

In order to verify that the observed trend with $Y_e$ is not an artifact introduced by our method of artificially varying the initial $Y_e$, we show the calculated Th/Eu ratios after 1~Gyr for all trajectories in our six hydrodynamical models in dependence of initial $Y_e$ and entropy $S$ in Figure~\ref{fig:ThEu_univ}. Again, a clear maximum at low entropies and around $Y_e = 0.15$ can be identified, confirming the results in Figure~\ref{fig:Th/Eu_trends}. The results also show that going towards higher $Y_e$, it is possible to maintain a large Th/Eu ratio as long as the tracer particle is ejected with a higher initial entropy. 




In the following, we discuss the main features of Figure~\ref{fig:Th/Eu_trends}: the (global) maximum in actinide production between $0.12 < Y_e < 0.17$, followed by a minimum at lower $Y_e$, and the position of the maximum which can be found at different $Y_e$ values for the different trajectories.

\section{Discussion}
\label{sec:discussion}
In section~\ref{sec:results} we showed the sensitivity of actinide yields on the nuclear physics side as well as the hydrodynamical conditions. As seen in Figures~\ref{fig:observ_ratios}~\&~\ref{fig:observ_ratios2}, the nuclear mass model has a large impact on the actinide production, and on the Th/Eu abundance ratio in particular (see also \citealt{holmbeck2019a}). Since the $\beta$-decay rates are unchanged between FRDM, ETFSI-Q, and DZ10, the difference comes from the different r-process reaction paths that are determined by the competition of neutron captures ((n,$\gamma$) reactions) and photodisintegrations (($\gamma$,n)). The relative strengths of neutron capture rates and photodisintegration rates is set by the reaction Q-values, i.e., the neutron separation energies, which in turn depend on the differences in nuclear masses. 

Just like (n,$\gamma$) and ($\gamma$,n) reactions, $\beta$-decays on the r-process path affect the final actinide abundances to a considerable degree, as is shown by model D3C* in Figure~\ref{fig:observ_ratios2}. The set of \cite{marketin2016} allows for first-forbidden transitions, which leads to faster decay rate predictions in the nuclei with large mass number $A$. This results in a speed-up of the reaction flow through the actinides and lower actinide abundances at any given time during the r-process calculation. This effect was also described in \cite{eichler2015}. In the following, we want to discuss the various factors that favour or  impede the Th production in greater detail.

\subsection{Why is Th most efficiently produced around $Y_e = 0.15$?}
\begin{figure}
  \includegraphics[width=\columnwidth]{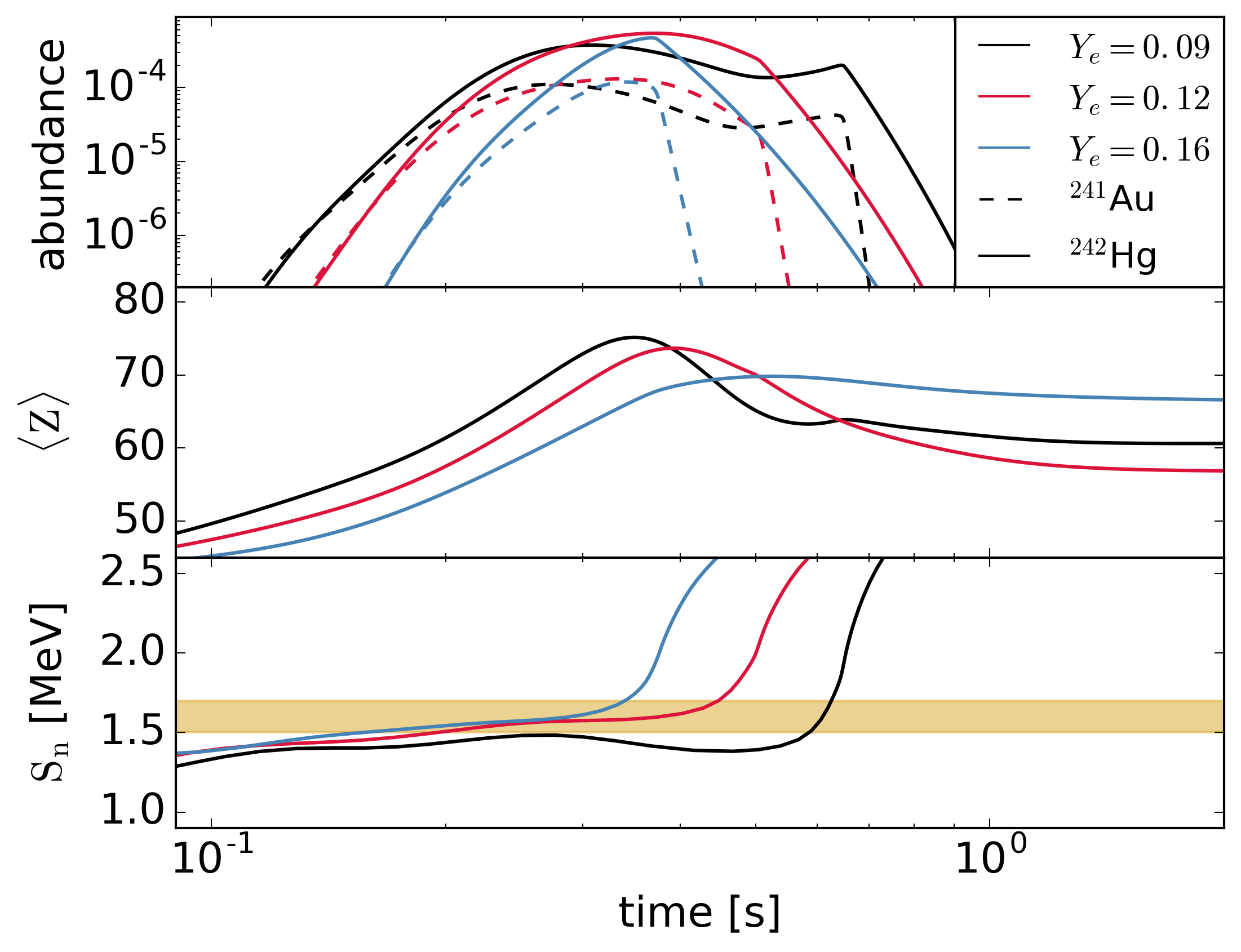}
  \caption{Comparison of three different initial $Y_e$ values on the example of R1010-rep. The $Y_e$ values of 0.16 and 0.09 lead to maxima in the final Th abundance, while $Y_e = 0.12$ represents a Th minimum (see Fig.~\ref{fig:Th/Eu_trends}, top left panel). Top: Evolution of $^{241}$Au and $^{242}$Hg abundances. Middle: average proton number $\left< Z \right>$. Bottom: average neutron separation energy $S_n$ of nuclei in the mass range $220 \leq A \leq 260$. \label{fig:freezeout}}
\end{figure}

In our calculations, the thorium abundance does not linearly depend on the initial neutron-richness (see Figure~\ref{fig:Th/Eu_trends}). Instead, most of the trajectories we have investigated produce the highest actinide yield around $Y_e = 0.16$, followed by a minimum (around $Y_e = 0.12$), and roughly constant abundances for even lower initial $Y_e$. Here we discuss the origin of these particular trends and why the exact location of the maximum and minimum Th abundance slightly vary for the individual trajectories. \cite{holmbeck2019a} have already demonstrated the dependence of the actinide and Eu abundances on the fission cycles. This effect is also apparent in our calculations and can be traced, for instance, by following the abundances of $^{241}$Au and $^{242}$Hg (i.e., the most abundant nuclei with relatively long half-lives at the $N=162$ neutron number and precursor nuclei for $^{232}$Th) and comparing their abundance evolution to the average proton number $\langle Z \rangle$ of the composition as an indicator of fission cycles. The top and middle panel of Figure~\ref{fig:freezeout} clearly show such a relation for the three runs in R1010 where the Th abundance reaches extreme values. The abundances of $^{241}$Au and $^{242}$Hg start to increase only after 0.1 to 0.2~s, when the reaction flow reaches the actinide region for the first time. Due to their relatively long half-lives, material starts to pile up in these two isotopes for the next 0.1 to 0.2~s. During this time the actinide abundances increase, while the abundances of seed nuclei with lower mass numbers gradually decline, leading to a slowdown in the production of new actinide nuclei. If the r-process freeze-out occurs around that time when the actinide abundance is highest (i.e., our case $Y_e = 0.16$ in Figure~\ref{fig:freezeout}), the resulting $^{232}$Th abundance will be very high. If, however, the r-process continues beyond that point, neutron captures carry away material from the $A=240$ region and into a part of the nuclear chart where fission occurs, thus destroying actinides and producing fission fragments around the second peak and/or rare-earth peak. This phase coincides with $\langle Z \rangle$ and the $^{241}$Au and $^{242}$Hg abundances decreasing in Figure~\ref{fig:freezeout}. If the r-process freezes out just after this first fission cycle has been completed (i.e. our case $Y_e = 0.12$), the composition decaying to stability will have lower actinide abundances and higher lanthanide abundances than the previously discussed case. In conditions with even higher neutron densities, the fission fragments of the first fission cycle continue capturing neutrons, which eventually leads to an increase in actinide abundances yet again (see our case $Y_e = 0.09$ in Figure~\ref{fig:freezeout}).

\begin{figure*}
  \includegraphics[width=0.5\textwidth]{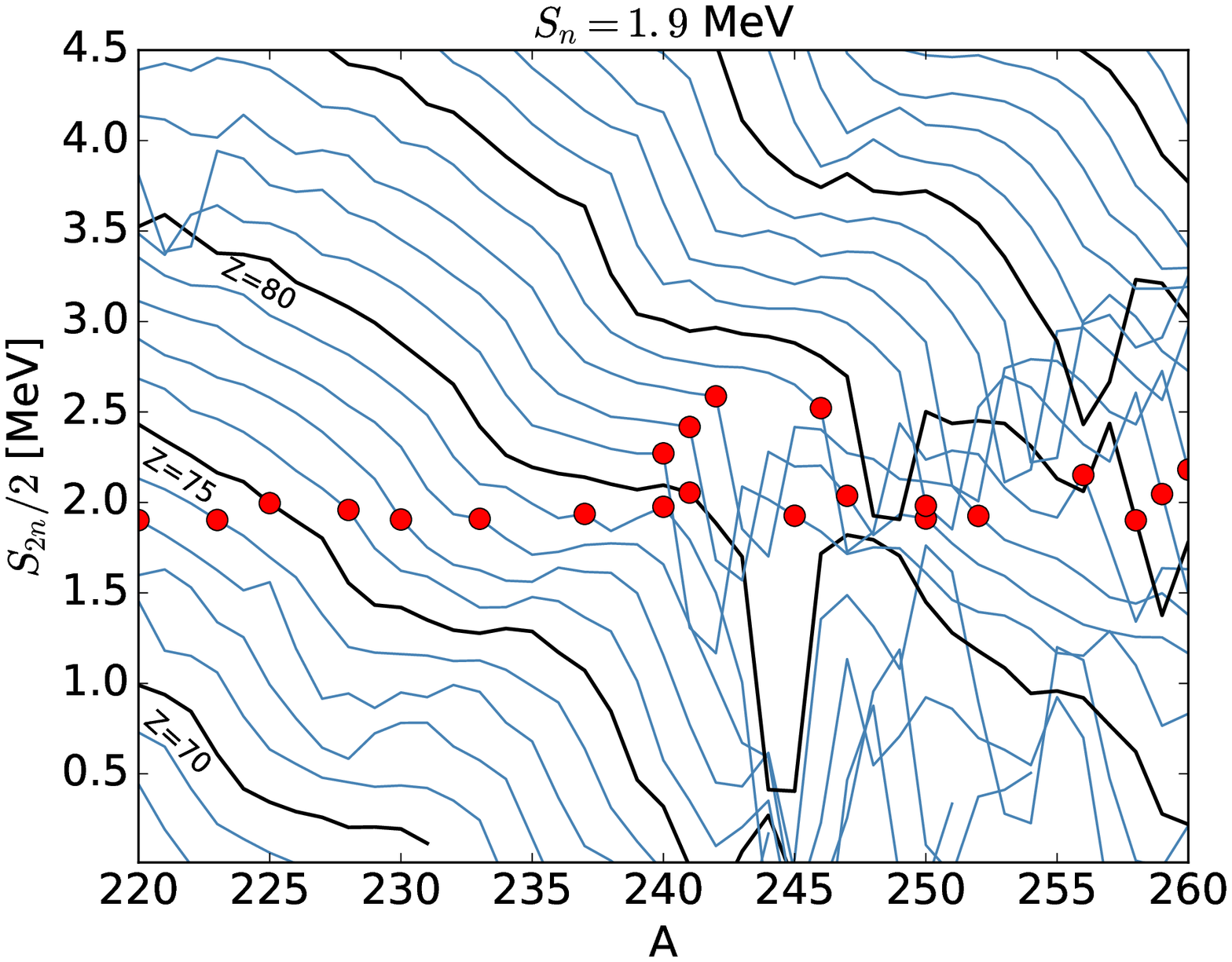}
  \includegraphics[width=0.5\textwidth]{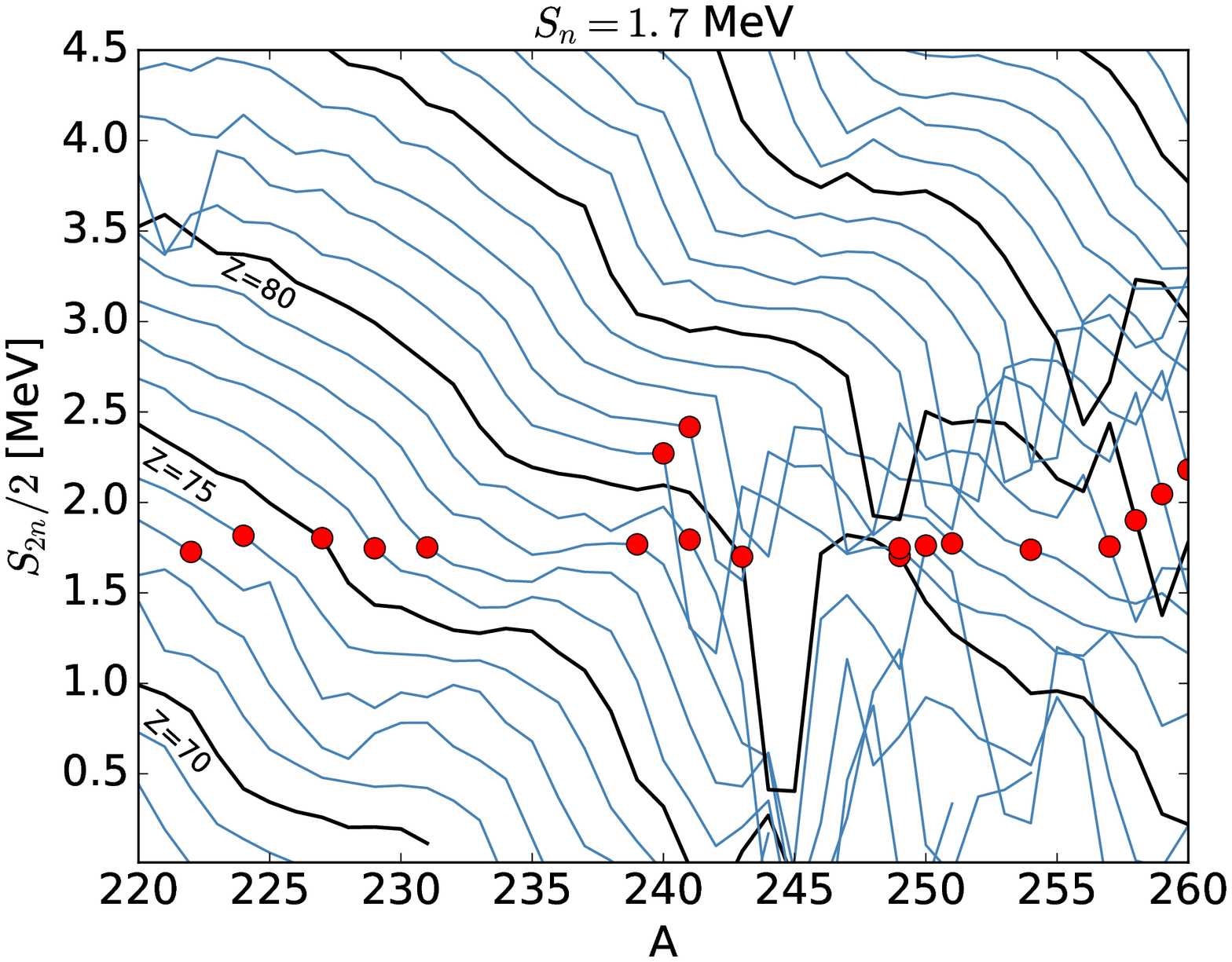} \\
  \includegraphics[width=0.5\textwidth]{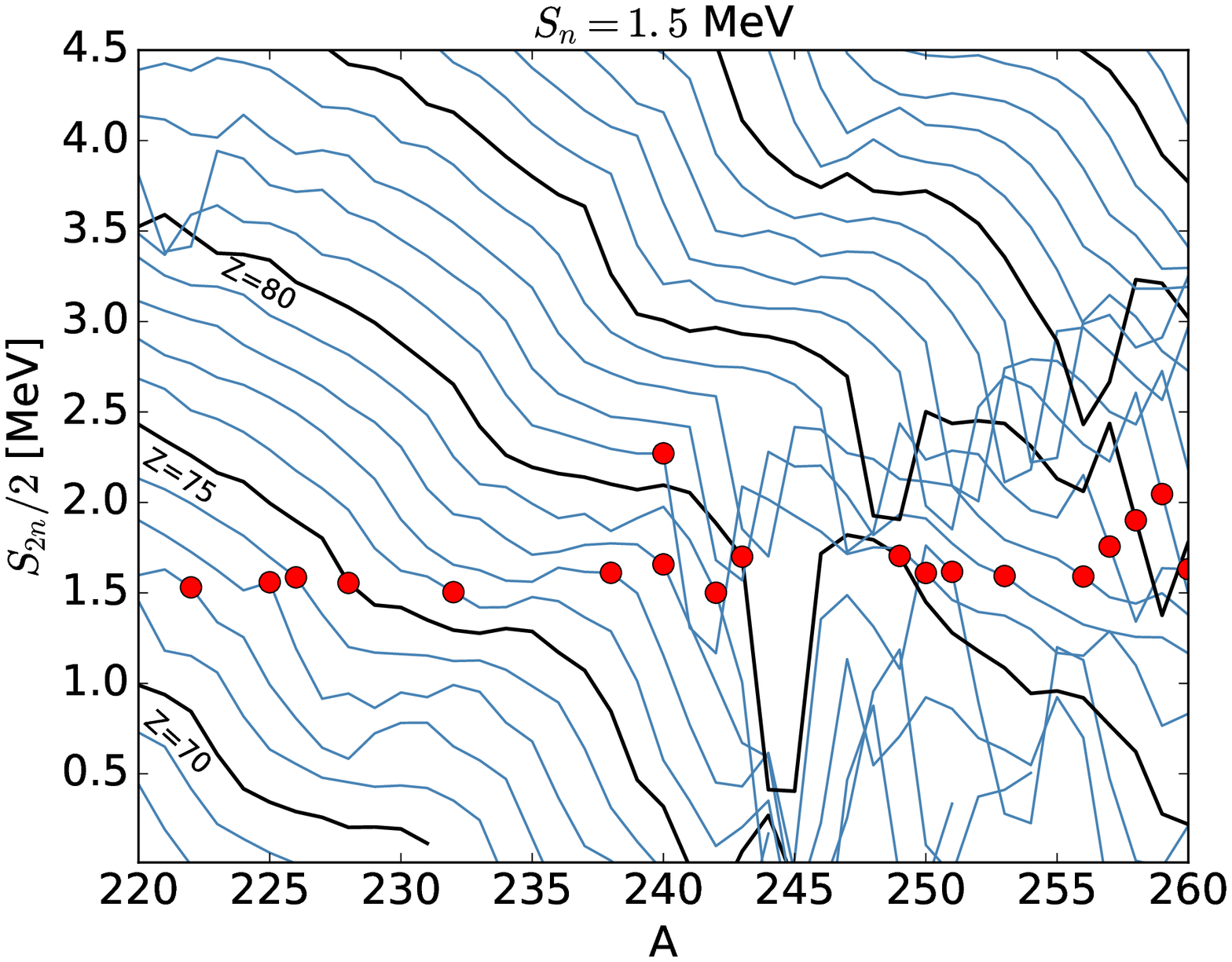}
  \includegraphics[width=0.5\textwidth]{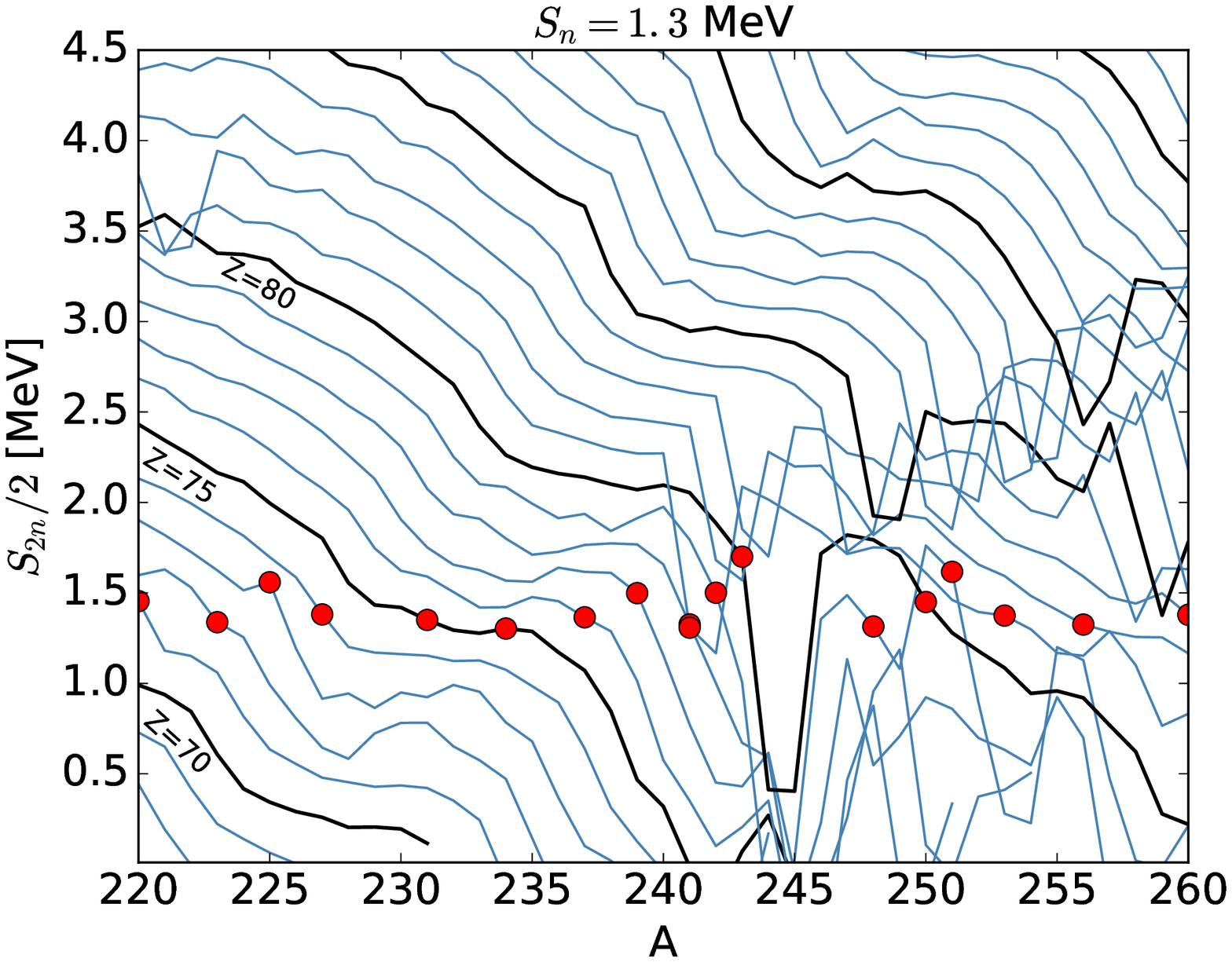}
  \caption{Two-neutron separation energies for the FRDM model, as a dependence of mass number. Equal proton numbers (i.e. isotopic chains) are connected by lines, with every fifth line in black. The red dots indicate typical r-process paths in (n,$\gamma$)-($\gamma$,n) equilibrium for different conditions. Top left to bottom right: $S_n = 1.9, 1.7, 1.5, 1.3$ MeV. \label{fig:s2n_path}}
\end{figure*}

However, we observe an additional effect that comes into play in the case of R1010-rep, where the variations in the individual abundances are strongest (see Figure~\ref{fig:Th/Eu_trends}). The bottom panel in Figure~\ref{fig:freezeout} shows the neutron separation energies at which nuclei are most abundant in (n,$\gamma$)-($\gamma$,n) equilibrium during the r-process. The shaded area highlights a range $1.5 \textrm{~MeV} < S_n < 1.7 \textrm{~MeV}$, where the reaction flow can easily bypass nuclei with mass numbers $A=232 - 238$ in the FRDM, as shown in the following. 
In hot r-process conditions, nuclei that are located on the reaction path can be identified based purely on their two-neutron separation energy, as well as the neutron density $n_n$, and the temperature $T$ (see, e.g., \citealt{thielemann2017}). The $S_{2n}/2$ values predicted by the FRDM mass model in the mass region $220<A<260$ are shown in Figure~\ref{fig:s2n_path}, together with typical r-process paths favouring nuclei around $S_{2n}/2 = 1.9, 1.7, 1.5, 1.3$~MeV (red dots). Every line represents an isotopic chain, with every fifth line drawn in black. Important to note is the saddle point structure for $Z=75-80$ at mass numbers $A=230-240$. If the freeze-out conditions favour nuclei close to $S_{2n}/2=1.7$~MeV, a gap opens at these mass numbers and the flow moves from $(Z,A)=(77,231)$ directly to $(Z,A)=(78,239)$, bridging all isotopes in between. The resulting low abundances of nuclei with mass numbers $A=232-240$ directly impedes the production of $^{232}$Th and its long-lived precursor isotopes, $^{236}$U and $^{240}$Pu (see section~\ref{sec:results}). If, however, the conditions at freeze-out favour nuclei with higher (i.e., closer to ``stability''; top left panel) or lower neutron separation energy (more neutron-rich; bottom right panel), the gap is closed and nuclei with $A=230-240$ are produced in larger amounts.

The lower panel of Fig~\ref{fig:freezeout} shows that the $Y_e = 0.09$ case freezes out in conditions with $S_n < 1.5$~MeV, thus enabling the efficient production of precursor nuclei for $^{232}$Th in the mass range $A=232 - 238$ and enhancing the effect of the fission cycles. 

\subsection{What determines the thorium and europium abundance trends in different $Y_e$ conditions?}
\label{sec:trends_analysis}
\begin{figure*}
  \centering
  \includegraphics[width=\textwidth]{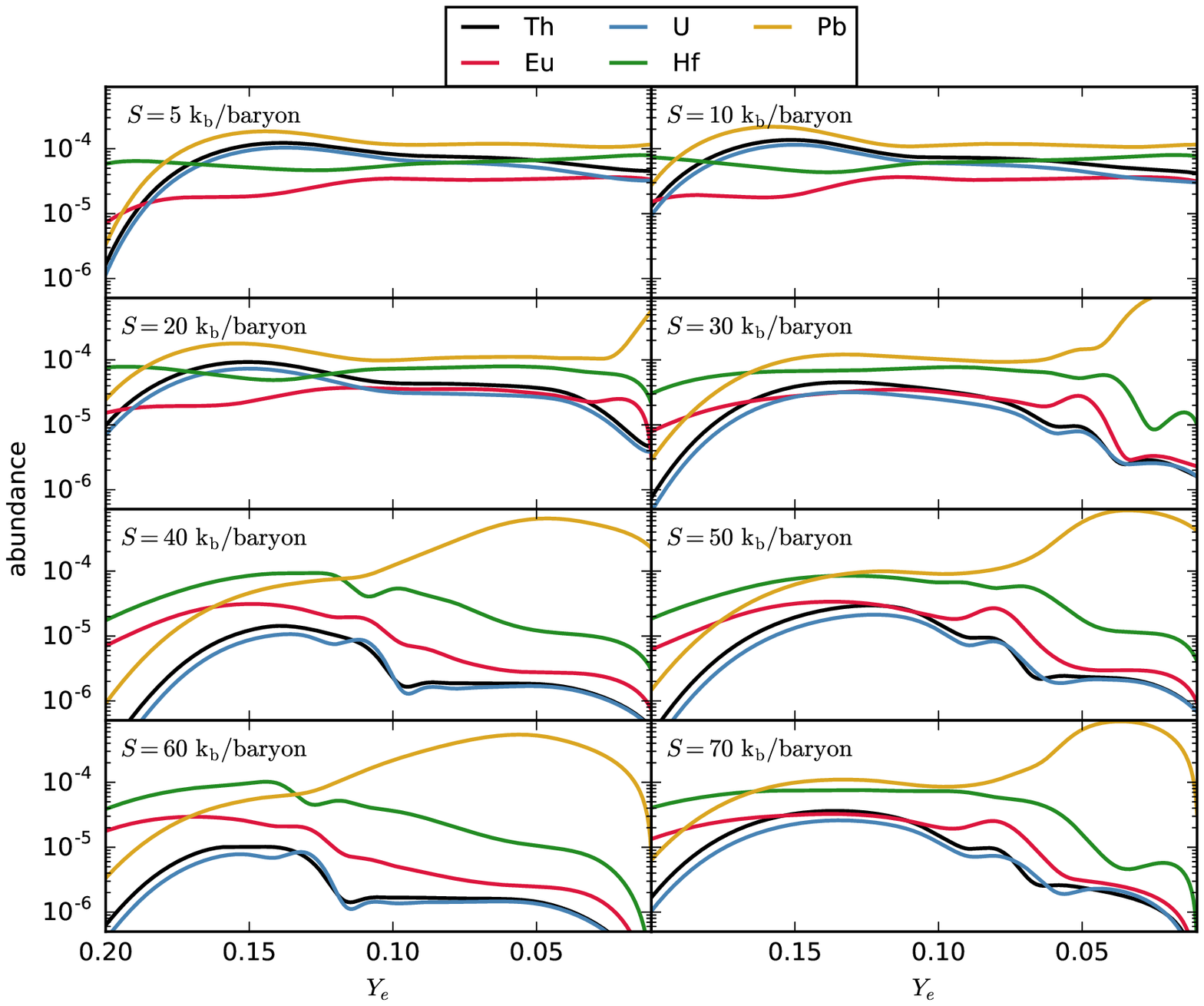}
  \caption{Same as Figure~\ref{fig:Th/Eu_trends}, but for trajectories with different initial entropies $S$. All trajectories are from Bs125 \citep{bovard2017}. \label{fig:varyS}}
\end{figure*}

The elemental abundances displayed in Figure~\ref{fig:Th/Eu_trends} show similar trends for most trajectories examined here. However, for a given electron fraction, the obtained abundances for the three elements (and their abundance ratios) are different for each case. Furthermore, the Bs125-rep case differs notably from the other cases. These observations reveal that the initial $Y_e$ does not solely determine the final Th and Eu abundances, and that other factors need to be taken into account. The Bs125-rep trajectory has the highest initial entropy (see Table~\ref{tab:trajectories}). We therefore test the impact of the initial entropy on the elemental abundances. In order to do this, we have picked from the simulation data of \cite{bovard2017} eight additional trajectories with starting entropies at $S=5, 10, 20, 30, 40, 50, 60, 70~k_B/\textrm{baryon}$, and calculated their Th, U, Pb, Eu, and Hf abundances for different initial $Y_e$ values. Figure~\ref{fig:varyS} shows that for entropies below $S = 20~k_b/\textrm{baryon}$, the abundance trends roughly follow the trends of the low-entropy cases of Figure~\ref{fig:Th/Eu_trends}, with a Th abundance maximum around $Y_e = 0.15$ coinciding with a local minimum in the Eu abundance. $S = 20~k_b/\textrm{baryon}$ also marks the point where the abundances for all three elements drop at the low-$Y_e$ end, since the high entropy counters the build-up of seed nuclei, resulting in lower r-process abundances for all mass ranges. For trajectories with $S > 30~k_b/\textrm{baryon}$, the Th/Eu abundance ratio is equal to unity or less for all initial $Y_e$ values. 
This means that while the Th and Eu abundance curves are following opposing trends for $S < 20~k_b/\textrm{baryon}$, they switch to a positive correlation for higher starting entropies. Furthermore, at any given initial electron fraction, $S < 20~k_b/\textrm{baryon}$ conditions lead to higher Th/Eu abundance ratios than conditions with higher entropies. Figure~\ref{fig:varyS} also reveals that ejecta with purely high-entropy conditions ( $S > 30~k_b/\textrm{baryon}$) are unlikely to produce large variations in the actinide-to-lanthanide abundance ratios.

\subsection{Nuclear Uncertainties}
\label{sec:discuss_nuc}
\begin{figure*}
\centering
\includegraphics[width=0.32\textwidth]{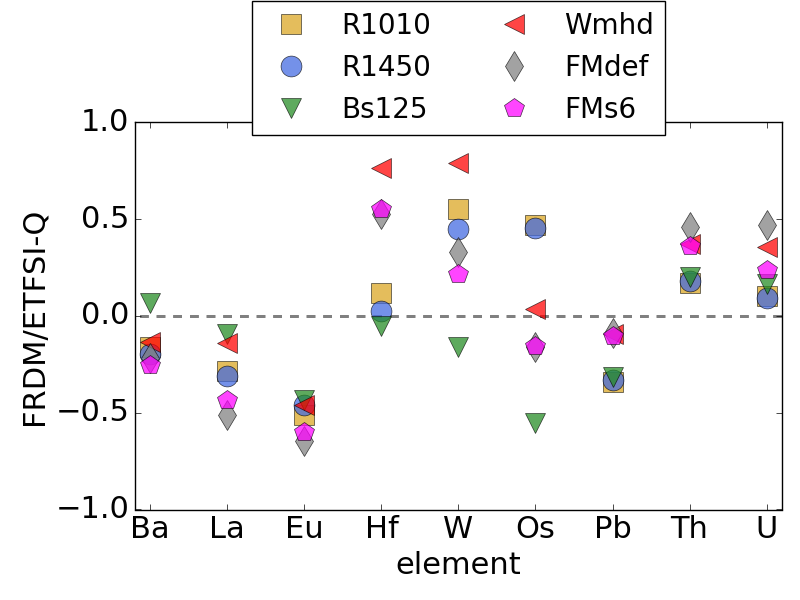}
\includegraphics[width=0.32\textwidth]{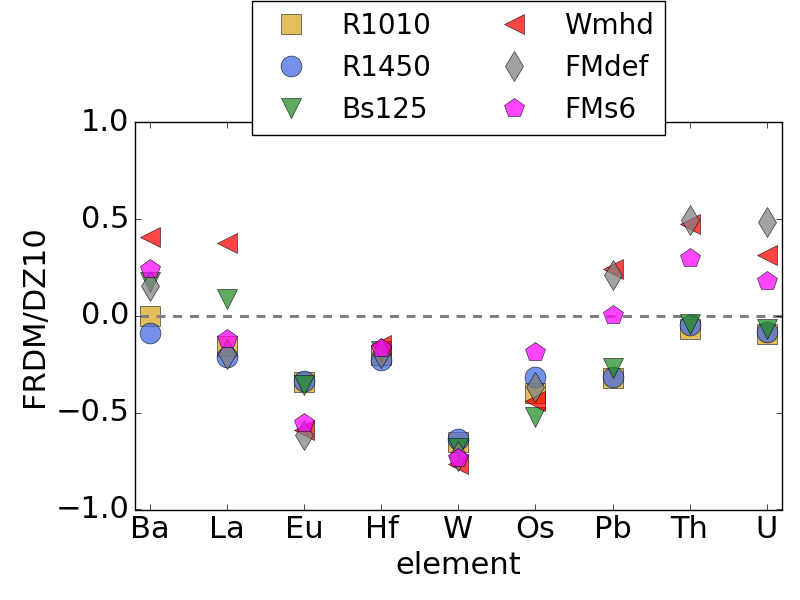}
\includegraphics[width=0.32\textwidth]{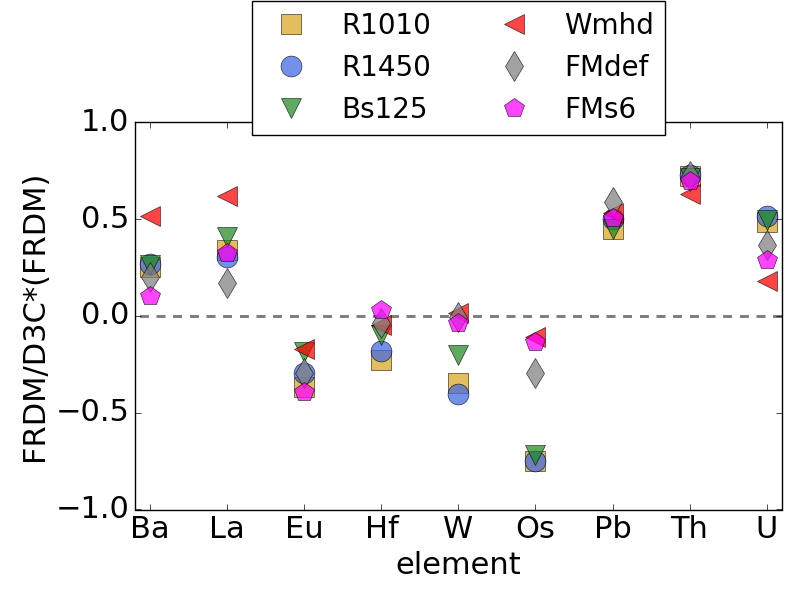} \\
\caption{Relative difference of elemental yields between FRDM and ETFSI-Q (left), FRDM and DZ10 (middle), and FRDM and D3C*(FRDM) (right) for the hydrodynamical models investigated. The values plotted correspond to $(Y_1-Y_2)/(Y_1+Y_2)$, where the indices $1$ and $2$ refer to the different mass models. \label{fig:massmodel_factors}}
\end{figure*}

The dependences of the actinide and europium abundances on the mass model, $\beta$-decays, and the fission fragment distribution have been discussed in \cite{goriely1999}, and, more recently on the example of hydrodynamical trajectories from NSM dynamical or disk ejecta, in \cite{wu2016,vassh2018,holmbeck2019a}. \cite{holmbeck2019a} found that in nucleosynthesis calculations using $\beta$-decay rates from \cite{marketin2016} (corresponding to our model D3C*(FRDM)), the variations in Th and Eu abundances with initial $Y_e$ are much smaller, and that the resulting Eu abundance is higher, while the Th yield is smaller. They relate these differences to two distinct regions in the nuclear chart where the \cite{marketin2016} rates are faster than the corresponding FRDM rates: (a) around $A=130$, carrying material away from the second peak and filling the rare-earth peak more efficiently, and (b) above $A = 190$, resulting in a similar effect, where nuclei are guided to the fissioning region more quickly, and therefore less material is stored in $\alpha$-decaying actinides after the freeze-out. In addition to these effects, we found that the $\beta$-decay half-life predictions of the nuclei on the $N=162$ isotone also play an important role. As mentioned before, $^{242}$Hg and $^{241}$Au are especially abundant during the r-process, and a strong connection can be observed between their abundances at the r-process freeze-out and the final abundance of $^{232}$Th (see Fig.~\ref{fig:freezeout}).
The half-lives of these two nuclei are not (yet) known experimentally, and the JINA reaclib $\beta$-decay rate \citep{moeller2003,mumpower2016} predicts a half-life of 0.03~s for $^{242}$Hg, while \cite{marketin2016} predict 0.001~s. For $^{241}$Au the predictions are 0.006~s \citep{moeller2003,mumpower2016} and 0.001~s \citep{marketin2016}, respectively. We test the impact of these half-live predictions on our results by exchanging the decay rate of $^{242}$Hg (along with its $\beta$-delayed neutron emission probabilities) in our nuclear reaction rate libraries FRDM and D3C*(FRDM) with the rate from the other set. Indeed, for R1010-rep, R1450-rep, FMdef-rep, FMs6-rep this leads to a change in the final $^{232}$Th abundance by factors between 1.49 and 2.25 (more Th whenever a \citealt{marketin2016} rate is replaced by an FRDM rate, and vice versa). In a second step, we exchange the rates of both $^{242}$Hg and $^{241}$Au simultaneously. This increases the effect slightly, with the range of factors changing to $1.63-2.49$. The original Wmhd-rep and Bs125-rep trajectories are only weakly affected by these changes, since the reaction paths in the relatively high-$Y_e$ (or high entropy in the case of Bs125) environments run closer to stability at the time of freeze-out, bypassing $^{242}$Hg and $^{241}$Au. To further quantify the impact of the chosen mass model, we show in Fig.~\ref{fig:massmodel_factors} the relative changes in elemental abundances for several chosen elements beyond the second r-process peak between ETFSI-Q (left panel), DZ10 (middle panel), and D3C*(FRDM) (right panel) with respect to FRDM. The colours and symbols represent the integrated ejecta yields for the different hydrodynamical models, and the values are the relative abundance differences, $(Y_1-Y_2)/(Y_1+Y_2)$, where the indices $1$ and $2$ refer to the different mass models. This enables an assessment of the uncertainties of the individual elements with respect to the nuclear mass model: an elemental yield is sensitive only to the mass model if the points lie far away from 0, but close to each other and sensitive to both the nuclear and hydrodynamical model if the spread of points is large. Although it is difficult to draw conclusions based on Fig.~\ref{fig:massmodel_factors} alone, the data suggest that the ejecta composition of Wmhd, FMdef, and FMs6 are generally more sensitive to the nuclear mass model (with respect to the elements shown here), while the more neutron-rich scenarios reveal a more robust behaviour (i.e., they are often closer to the equality line). Furthermore, Th and U have the almost identical dependencies on both the nuclear and the hydrodynamical models, suggesting once more that U/Th is the most reliable chronometer pair at present.

\section{Possible observables connected to high actinide yields}
\label{sec:observables}
Figure~\ref{fig:Th/Eu_trends} suggests that if actinide-boost stars originate from the same astrophysical site as ``regular'' r-II stars, they would have inherited elemental compositions from ejecta close to the $Y_e$ value where Th production is most efficient. For the same conditions, most hydrodynamical models predict a lower Eu abundance in comparison, since the Eu abundance is anti-correlated to Th for all models except for Bs125-rep. On the other hand, U correlates with the Th trend for all trajectories in Figure~\ref{fig:Th/Eu_trends}. In this section we want to identify other elements whose abundances in different $Y_e$ environments follow a specific evolution, and whether they follow the Eu trend, the Th trend, or whether there are elements that have their own individual behaviour. For this reason we calculate, for all stable elements starting at the second r-process peak ($Z=48$) and for our representative trajectories summarized in table~\ref{tab:trajectories}, the Spearman correlation coefficient $\rho$. 

\begin{figure*}
  \includegraphics[width=\textwidth]{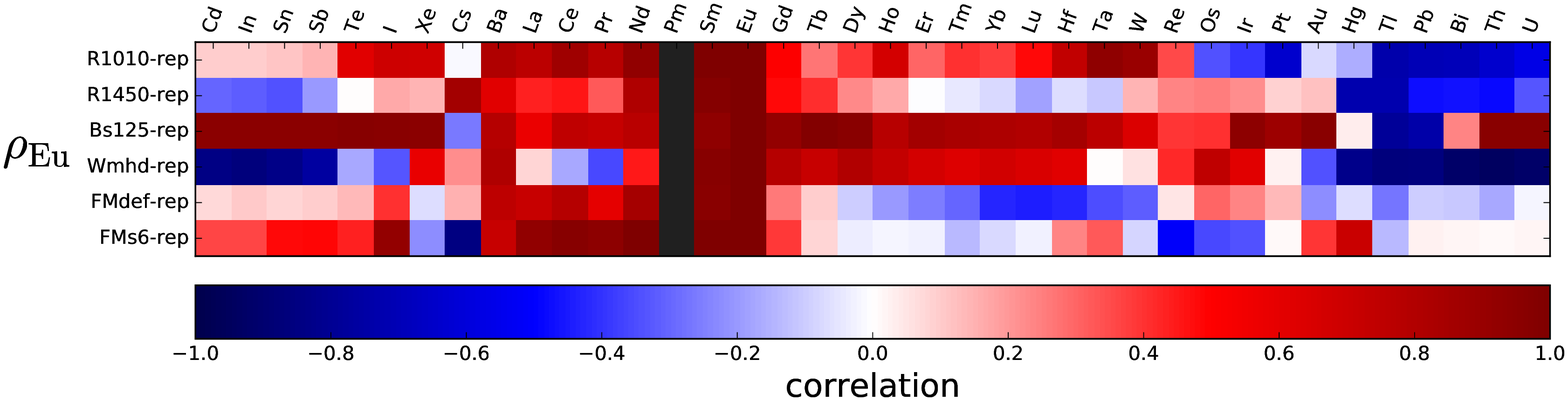} \\
  \includegraphics[width=\textwidth]{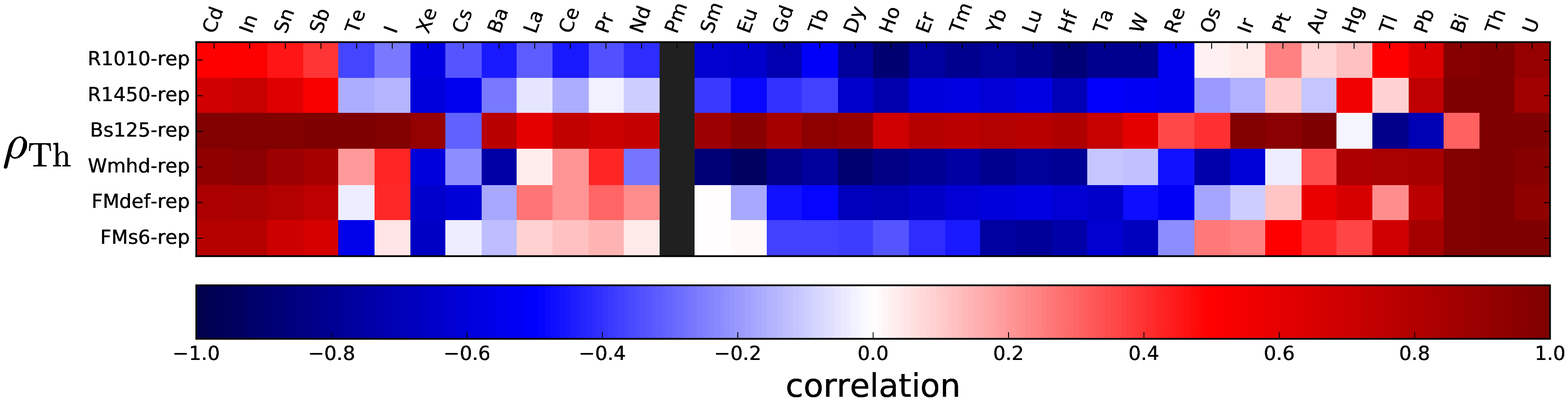} \\
  \caption{Spearman correlations with Eu (top) and Th (bottom) for all elements from Cd to U, using FRDM and based on our six representative trajectories and varying the initial electron fraction in the range $0.20 \geq Y_e \geq 0.01$ (see Fig.~\ref{fig:Th/Eu_trends}). \label{fig:correlations}}
\end{figure*}

Since we are interested in possible (anti-)correlations with Eu and Th, we calculate the correlation coefficient both with respect to Eu and to Th. The results are shown in Fig.~\ref{fig:correlations} for all stable or long-lived elements from $Z=48$ (Cd) to $Z=92$ (U). A value of $\rho_{\rm Eu,Th} = +1$ means perfect correlation, while $\rho_{\rm Eu,Th} = -1$ indicates perfect anti-correlation.

Fig.~\ref{fig:correlations} allows us to classify elements into Eu-like, Th-like, and independent elements, for a given hydrodynamical trajectory. For all trajectories used here, Ba, Nd, Sm, and Gd are Eu-like. Heavier lanthanides also generally follow the Eu trend, except for FMdef-rep and, to a certain extent, FMs6-rep and R1450-rep. Th-like elements for all conditions are the direct decay products of lighter fissioning nuclei and $\alpha$-decaying actinides, i.e., second peak elements Cd, In, Sn, and Sb, and heavier elements such as Pb, Bi, and U, while the third-peak elements Pt, Au, Hg, and Tl are also Th-like for most sites. As for anti-correlations, Tl and Pb seem to be anti-Eu for all models except Bs125-rep, while Cs and Re are always anti-Th. While the upper panel in Fig.~\ref{fig:correlations} shows that the disk trajectories have a unique pattern, with the second peak elements behaving as Eu and the heavier lanthanides anti-correlated, in the lower panel Bs125-rep can be identified as a special case, with all lanthanides correlating with both Th and Eu, while Cd, Cs, Hg, Tl, and Pb all strongly anti-correlate. This can be explained by the higher entropy of Bs125-rep (see section~\ref{sec:trends_analysis}). If elemental abundances of actinide-boost stars can be extracted with high enough precision and for a large sample of elements, the correlations discussed here could potentially be used to decide on the conditions in the ejecta that lead to the actinide-enhanced r-process composition. For instance, if actinide-boost stars showed no systematic irregularities with respect to Eu in the region from Te to Re, conditions with very low entropy and fast expansion velocities seem the best candidate (our model R1010-rep). If they had (systematically) larger Ce/Eu and Pr/Eu ratios (in addition to the higher Th/Eu), conditions with higher entropies and longer dynamical timescales would be able to explain that (like our models Bs125-rep, FMdisk-rep, and Wmhd-rep). Note that the trends obtained here can depend on the fission fragment distribution model employed. Some fission fragment distribution models predict strongly asymmetric fragment production also for light fissioning nuclei. These models could favour more positive correlations between lighter lanthanides (Ba to Eu) and Th.
The correlations also depend on the nuclear physics input. We show the correlations obtained with the other mass models and with the D3C* $\beta$-decay predictions in appendix~\ref{app:a}. Although there are differences for individual elements, the results are qualitatively the same.

\begin{figure*}
  \centering
  \includegraphics[width=0.49\textwidth]{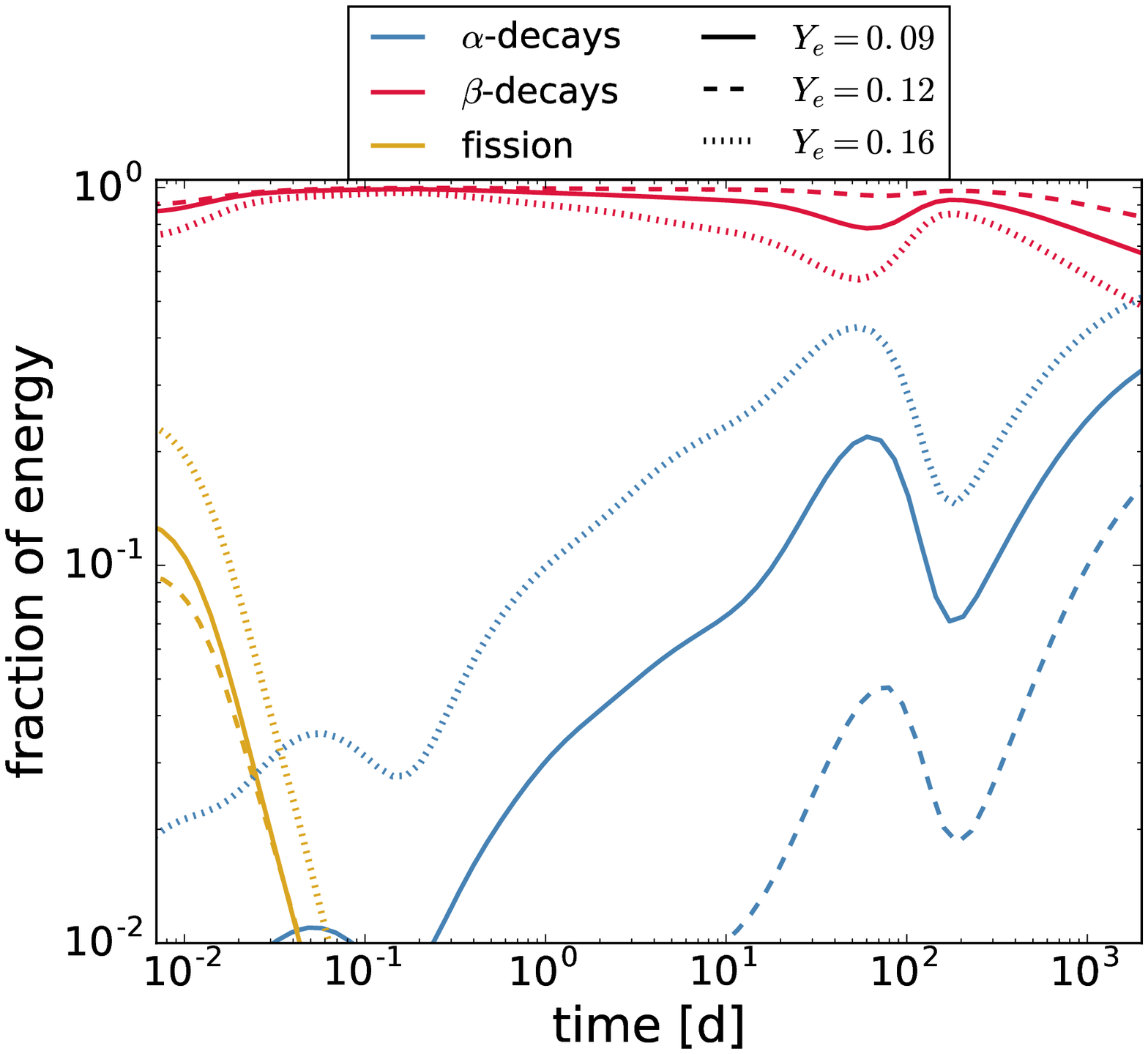}
  \includegraphics[width=0.49\textwidth]{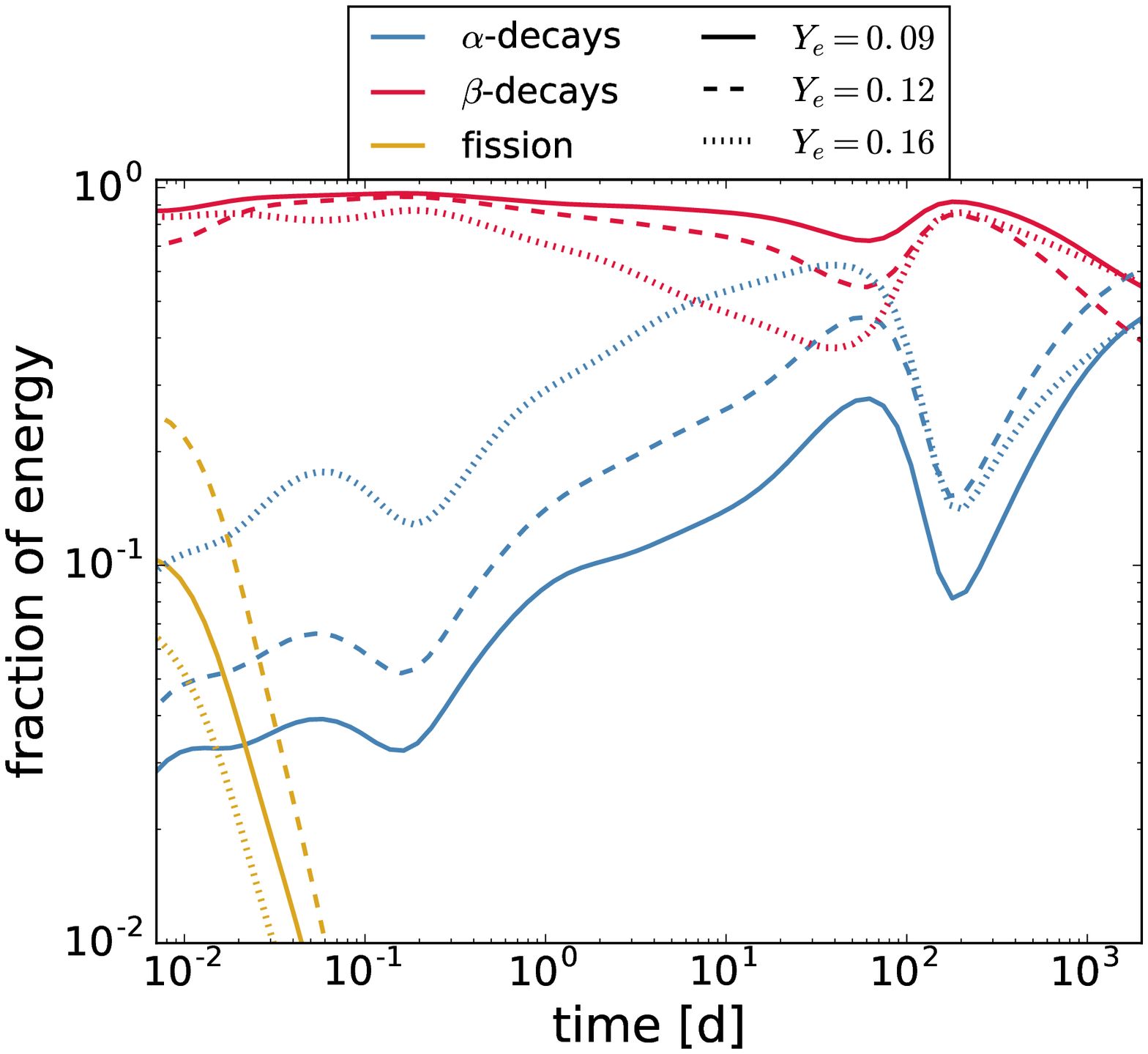} \\
  \caption{Fractions of total energy generation due to radioactive decays of r-process nuclei for $\alpha$-decays (blue), $\beta$-decays (red), and fission (green) on the example of the R1010-rep trajectory with different initial $Y_e$ values. Left: FRDM. Right: Duflo-Zuker. \label{fig:heating_contrib}}
\end{figure*}

So far, we have discussed possible (as of yet undetected) irregular signatures in the abundance patterns of actinide-boost stars. Another exciting possibility has been enabled only recently by the emergence of multi-messenger astronomy and the first detection of a binary neutron star merger, along with an electromagnetic afterglow powered by the decays of r-process nuclei (macro- or kilonova).
\cite{zhu2018} have identified $^{254}$Cf as a possible isotope whose half-life of $60.5 \pm 0.2$ days makes it a good candidate to find a signature of its decay in the kilonova light curve. \cite{wu2019} discuss the role of $\alpha$-decays in the late-time light curves of kilonovae. Both \cite{wu2019} and \cite{wanajo2018} point out the potential importance of the $^{72}$Zn~$\rightarrow ^{72}$Ga~$\rightarrow ^{72}$Ge decay chain. Variations in initial $Y_e$ in our dynamical ejecta models also have a significant effect on the nuclear heating rates in the ejecta at late times. As an example, the contributions from $\alpha$- and $\beta$-decays as well as spontaneous fission to the overall nuclear heating rate are shown in Figure~\ref{fig:heating_contrib} as a function of time for different initial $Y_e$ values for a trajectory from \cite{rosswog2013} (i.e., with very low entropy and rapidly expanding). In our calculations, the production of $^{254}$Cf is blocked by $\beta$-delayed fission of $^{254}$Am (and $^{254}$Bk in the case of FRDM; \citealt{panov2005}). However, we observe a contribution from $\alpha$-decays around 50~days that is caused by a decay chain starting at $^{225}$Ac all the way down to $^{209}$Bi (also visible in calculations from \citealt{barnes2016}~\&~\citealt{wanajo2018}). The strength of the $\alpha$-decay peak strongly correlates with the Th production, with the strongest contribution at $Y_e = 0.15-0.20$. The trends discussed here are similar for trajectories from \cite{bovard2017} with higher entropies, although the $Y_e=0.20$ calculation leads to a smaller $\alpha$-decay contribution.
Of course, the individual heating contributions also depend heavily on the adopted nuclear mass model, as was shown by \cite{barnes2016} and \cite{rosswog2017b}.

\section{Conclusions}
\label{sec:conclusions}
The varying Th/Eu abundance ratios in extremely metal-poor stars are in contrast to the observed robustness of the r-process abundance pattern for other elemental pairs. This raises the question whether the r-process compositions of actinide-boost stars originate from a different r-process site than the normal r-enhanced EMP stars or whether they inherited an r-process composition produced under slightly different conditions, but from the same site. The large range of calculated Th/Eu abundance ratios in Figures~\ref{fig:observ_ratios}~\&~\ref{fig:observ_ratios2} demonstrates that the Th/Eu ratio is heavily dependent on both the hydrodynamical conditions and the nuclear mass model. Furthermore, it can also be seen that all observed stellar Th/Eu abundance ratios could for instance be achieved by mixing different fractions of dynamical ejecta and disk ejecta in a single neutron star merger event. It would however also be possible for ejecta from accretion disks or MHD SNe to produce actinide-enhanced compositions, if they have predominantly low entropies and $Y_e = 0.1-0.15$ (see Fig.~\ref{fig:ThEu_univ}). For the pairs Pb/Th and U/Th the observed spread cannot be explained that easily by our models, since these elements always seem to be co-produced in similar amounts in our r-process models (except for artificial conditions with high entropy and very low $Y_e$, see the Bs125-rep case in Fig.~\ref{fig:Th/Eu_trends}). However, the observational sample is very small for these element pairs and most stars with measured Pb have log~$\epsilon$~(La/Eu)~$>0.25$, suggesting that they are not pure r-process stars. Future observations will help constraining the elemental ratios that need to be explained by the r-process alone.

In our calculations in which the conditions are neutron-rich enough for the r-process to reach the $N=162$ isotopes, the theoretical $\beta$-decay half-life predictions of $^{241}$Au and $^{242}$Hg play an important role in determining the final $^{232}$Th abundance, as these two nuclei are relatively long-lived in comparison to other nuclei on the r-process path in this mass region. Especially for $^{242}$Hg, the predictions from our two sets of theoretical $\beta$-decay predictions are a factor of 30 apart, leading to a difference by a factor of 1.5 to more than 2 in the final thorium yields, depending on the initial conditions.

A maximum in Th/Eu abundance ratios is reached when the initial neutron-to-seed ratio is sufficient to produce actinides in large amounts, but not high enough to drive a complete fission cycle. In slightly more neutron-rich conditions, most actinides undergo fission before the r-process freezes out, and the resulting composition is rich in lanthanides, but has relatively small actinide abundances. For even more neutron-rich conditions, fission cycles lead to elemental abundances at roughly constant values. This is not the case for conditions with large initial entropies, where very neutron-rich conditions result in unusual abundance patterns, with the third peak mainly composed of Pb isotopes. Keep in mind, however, that in numerical simulations the amount of matter ejected with these conditions is very small compared to the total ejecta mass.

At low entropies, a small difference in $Y_e$ can translate into a large difference in the Th/Eu abundance ratio and therefore a single site (including dynamical and disk ejecta mixed always in similar ratios for all NSMs) with slightly varying conditions could also be responsible for the observed spread in Th/Eu.

Apart from this open question, we find that actinide-boosted compositions also exhibit peculiar abundances in other (lighter) elements (as shown in Fig.~\ref{fig:correlations}). A larger sample of detected kilonovae in the future will show whether the actinide content in the red component can vary or not, i.e., whether the conditions in neutron-rich NSM ejecta are variable or not (Fig.~\ref{fig:heating_contrib}). With D3C*(FRDM), the variations in Th/Eu are very small for different $Y_e$ values in one single trajectory. The differences between the hydrodynamical models with this mass model seen in Fig.~\ref{fig:observ_ratios2} mainly arise from the different hydrodynamical conditions encountered.

Although our results leave open the question of the origin of actinide-boost stars, this work has advanced our understanding of how variations in actinide abundances can arise. Future experimental and theoretical improvements are necessary to constrain masses and $\beta$-decay half-lives of neutron-rich nuclei, especially around the $N~=~162$ neutron number.

\vspace{3cm}
\section*{Acknowledgments}
We are grateful to Luke Bovard, Rodrigo Fern\'andez, Roger K\"appeli, Luciano Rezzolla, and Stephan Rosswog for their valuable input. Furthermore, we would like to thank Anna Frebel, Camilla Juul Hansen, Terese Hansen, Alex Ji, Oleg Korobkin, Friedrich-Karl Thielemann, and Meng-Ru Wu for inspiring discussions. M. E. was supported by the Swiss National Foundation under grant no. P2BSP2\_172068. T.R. acknowledges support by the COST Action CA16117 (ChETEC). This work was funded by Deutsche Forschungsgemeinschaft through SFB 1245, and by the ERC grant 677912 EUROPIUM.


\appendix
\section{Elemental correlations with other reaction rate libraries}
\label{app:a}
In Figure~\ref{fig:correlations} we have shown which elemental abundances correlate with Eu and Th when the initial $Y_e$ of our representative trajectories is varied, for the case of the default FRDM mass model. Here we show the results obtained with the other three reaction rate libraries used in this study (see section~\ref{sec:method}). Figures~\ref{fig:corr_E},~\ref{fig:corr_D},~\&~\ref{fig:corr_M} show the results for ETFSI-Q, Duflo-Zuker, and D3C*(FRDM), respectively.

\begin{center}
\begin{figure}[b]
  \includegraphics[width=0.85\columnwidth]{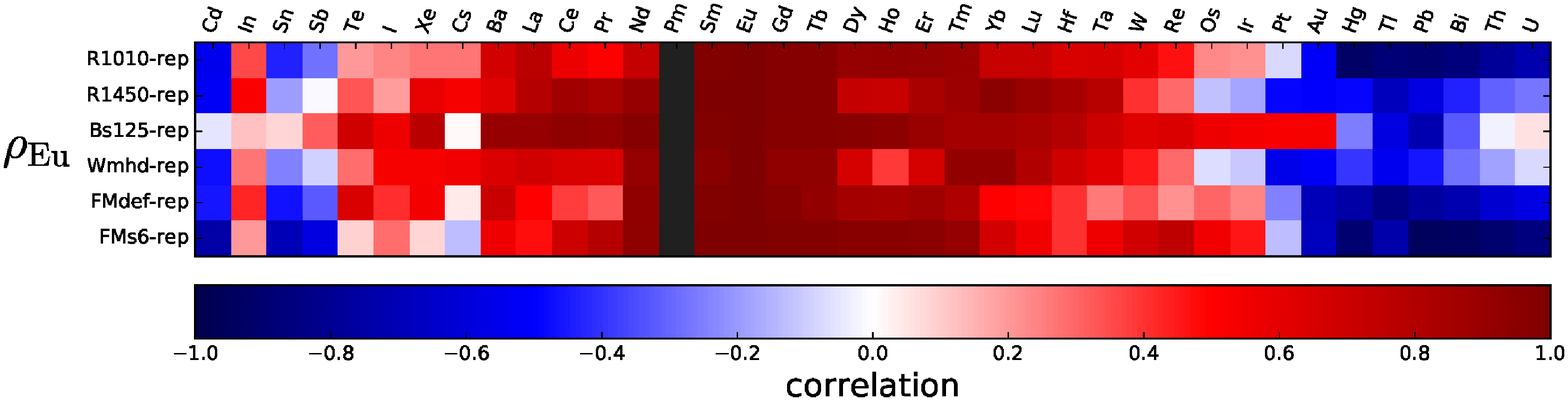} \\
  \includegraphics[width=0.85\columnwidth]{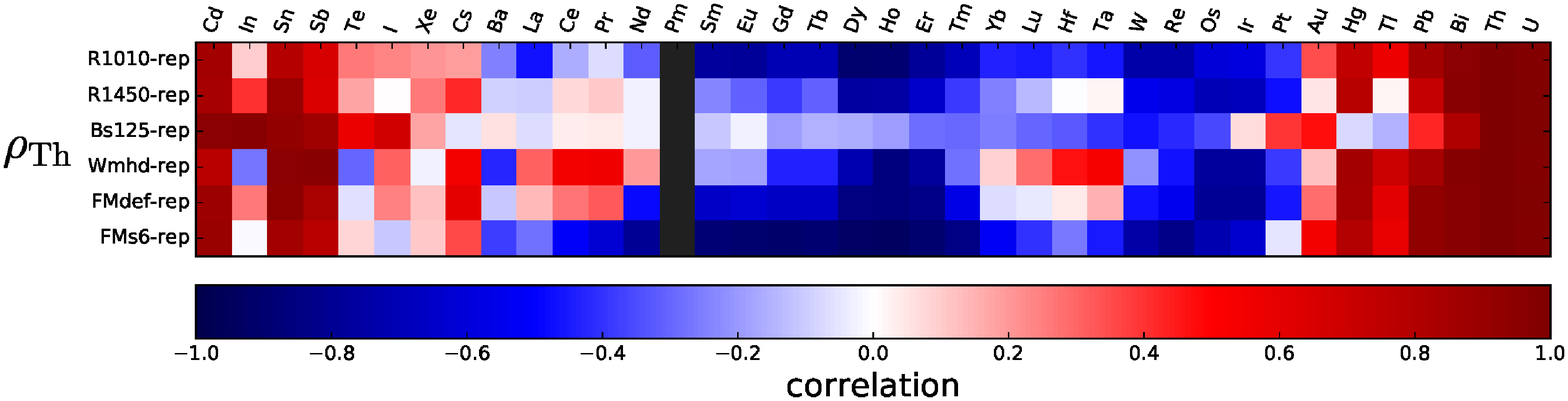}
  \caption{Same as figure~\ref{fig:correlations}, but for ETFSI-Q. \label{fig:corr_E}}
\end{figure}

\begin{figure}[t]
  \includegraphics[width=0.85\columnwidth]{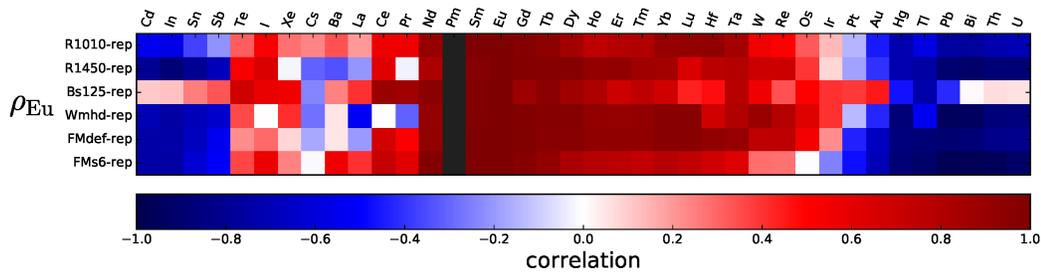} \\
  \includegraphics[width=0.85\columnwidth]{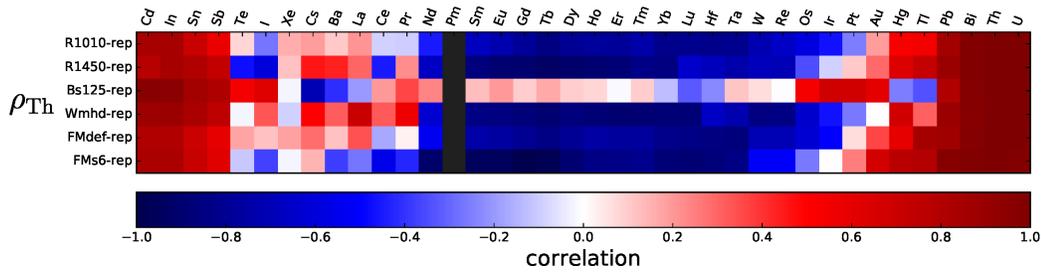}
  \caption{Same as figure~\ref{fig:correlations}, but for Duflo-Zuker. \label{fig:corr_D}}
\end{figure}

\begin{figure}[b]
  \includegraphics[width=0.85\columnwidth]{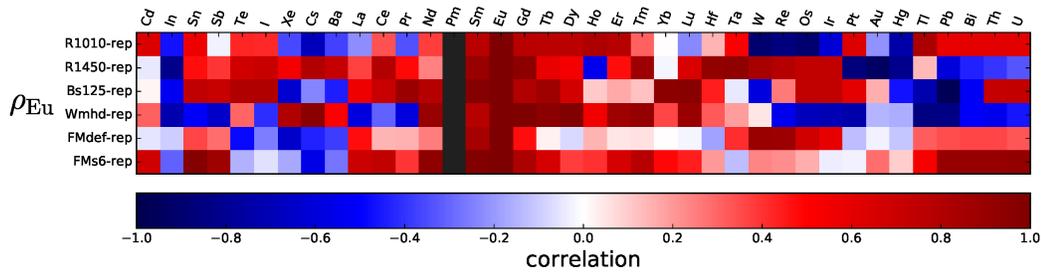} \\
  \includegraphics[width=0.85\columnwidth]{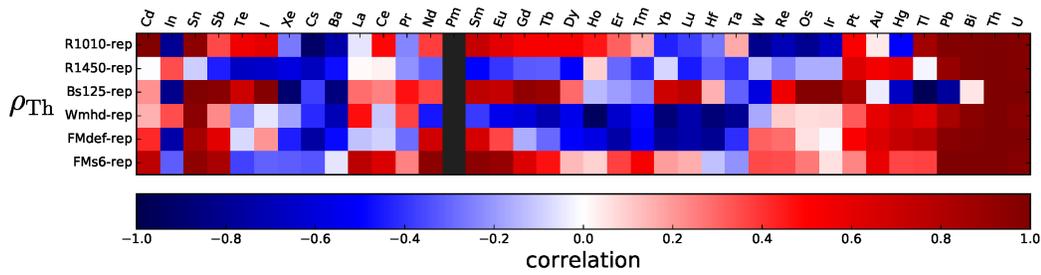}
  \caption{Same as figure~\ref{fig:correlations}, but for D3C*(FRDM). \label{fig:corr_M}}
\end{figure}

\end{center}

\newpage

\bibliographystyle{yahapj}
\bibliography{Masterbib_2018}

\end{document}